\documentclass[%
 reprint,
nofootinbib,
 amsmath,amssymb,
 aps,
]{revtex4-1}

\usepackage{graphicx}
\usepackage{dcolumn}
\usepackage{bm}
\usepackage{caption}
\usepackage{subcaption}
\usepackage{appendix}

\begin{document}

\preprint{APS/123-QED}

\title{Design Considerations of Gold Nanoantenna Dimers for Plasmomechanical Transduction}

\author{Zubair Buch}
\affiliation{Institute of Sensor and Actuator Systems, TU Wien, 1040 Vienna, Austria.}
\author{Silvan Schmid}%
 \email{silvan.schmid@tuwien.ac.at}
\affiliation{Institute of Sensor and Actuator Systems, TU Wien, 1040 Vienna, Austria.}

\date{\today}

\begin{abstract}
Internal optical forces emerging from plasmonic interactions in gold nanodisc, nanocube and nanobar dimers were studied by finite element method. A direct correlation between the electric-field enhancement and optical forces was found by observing largest magnitude of optical forces in nanocube dimers. Moreover, further amplification of optical forces was achieved by employing optical power of the excitation source. The strength of optical forces was observed to be governed by the magnitude of polarisation density on the nanoparticles, which can be varied by modifying the nanoparticle geometry and source wavelength. This study allows us to recognise that nanoparticle geometry along with the inter-dimer distance are the most prominent design considerations for optimising optical forces in plasmonic dimers. The findings facilitate the realisation of all optical modulation in a plasmomechanical nanopillar system, which has promising applications in ultra-sensitive nanomechanical sensing and building reconfigurable metamaterials.

\begin{description}
\item[Keywords]
optical forces, nano-dimers, plasmomechanics, FEM simulations
\end{description}
\end{abstract}

\maketitle


\section{INTRODUCTION}

The property of noble metal nanostructures to exhibit surface plasmon resonance, the collective oscillation of the electron cloud on the metal surface in response to the oscillating electric field of the incoming light, avails them novel applications for sensing and detection at the nanoscale.  The optical properties of metal nanoparticles [MNPs] have been explored for enhancing different processes such as surface enhanced raman spectroscopy \cite{Reguera2017}, metal enhanced fluorescence \cite{Fothergill}, near-field microscopy \cite{Uebel} etc. Moreover, the interaction of MNPs with light leads to the polarisation of electric charge on the nanoparticle surface, enabling it to induce an electrostatic force on a nearby charged nanoparticle. Depending on the nature of the polarisation induced, MNPs can exert an attractive or repulsive force on adjacent nanoparticle(s) in vicinity. The potential of these electrostatic forces, also termed as optical forces has been explored for various applications such as optical manipulation \cite{Kingsley}, optical tweezing \cite{Bustamante}, and optical trapping of molecules and other nanoscale objects \cite{Ashkin}. In most cases for optical trapping of molecules and smaller entities the excitation condition demands for a tightly focused laser beam since the distribution of optical forces due to field gradient is limited to the diffraction limit of light. This often sets a bottleneck on the sizes of the particle that can be trapped optically. MNPs can address this limitation due to their ability to confine the incoming light below its diffraction limit through local field enhancement of the electric field \cite{Raziman}. MNPs strongly enhance the incoming field around their edges and in the gap regions for an ensemble of nanoparticles, avoiding rigorous procedures carried for optical trapping such as beam steering. 

Optical forces realised in MNPs are usually very small in magnitude to realise any motion in bulk systems. However, the opto-mechanical force balance at the nanoscale allows unique opportunities for leveraging these forces to drive the mechanics in nanomechanical systems. The integration of optical properties of MNPs with the mechanical properties of nanomechanical resonators opens up new possibilities for optically designing and controlling the motion in these hybrid systems, while at the same time benefiting from the miniaturization for their effortless integration with next-generation technological platforms. So far the utilisation of optical forces has mostly been implemented in studies focused on the trapping of molecules and similar entities. However, their potential for independently driving motion at the nanoscale has not been well explored. Here, we present a numerical study for the case of a plasmomechanical nanopillar-antenna system, which consists of a singly clamped nano-pillar dimer structure with metal nanoparticle on top of it, and report on design considerations of metal nano-antennas with the aim of maximising electrostatic forces between nanoparticles in a dimer for prospects of optically driving motion in these systems. The proposed nano-pillar antenna configuration offers the possibility to combine the free-space adressability of metal nanoparticles with three dimensional motion of nano-pillar resonators, which avails this system advantages over other nanomechanical structures such as beams, bridges or strings. The actuation and transduction of nanomechanical resonators by optical forces generated in MNPs can be utilised for multiple purposes such as ultra-sensitive force and mass sensing applications \cite{Rossi}, optical modulation and reconfigurability \cite{Song} etc. A recent study reported the generation of optical forces in a lateral configuration in gold nanorod based system, and consolidated the key experimental demonstration of motion driven at the nanoscale by optical forces originating from plasmonic interactions between the nanorods \cite{Tanaka}.

\begin{figure}
    \centering
    \includegraphics[width=0.43\textwidth]{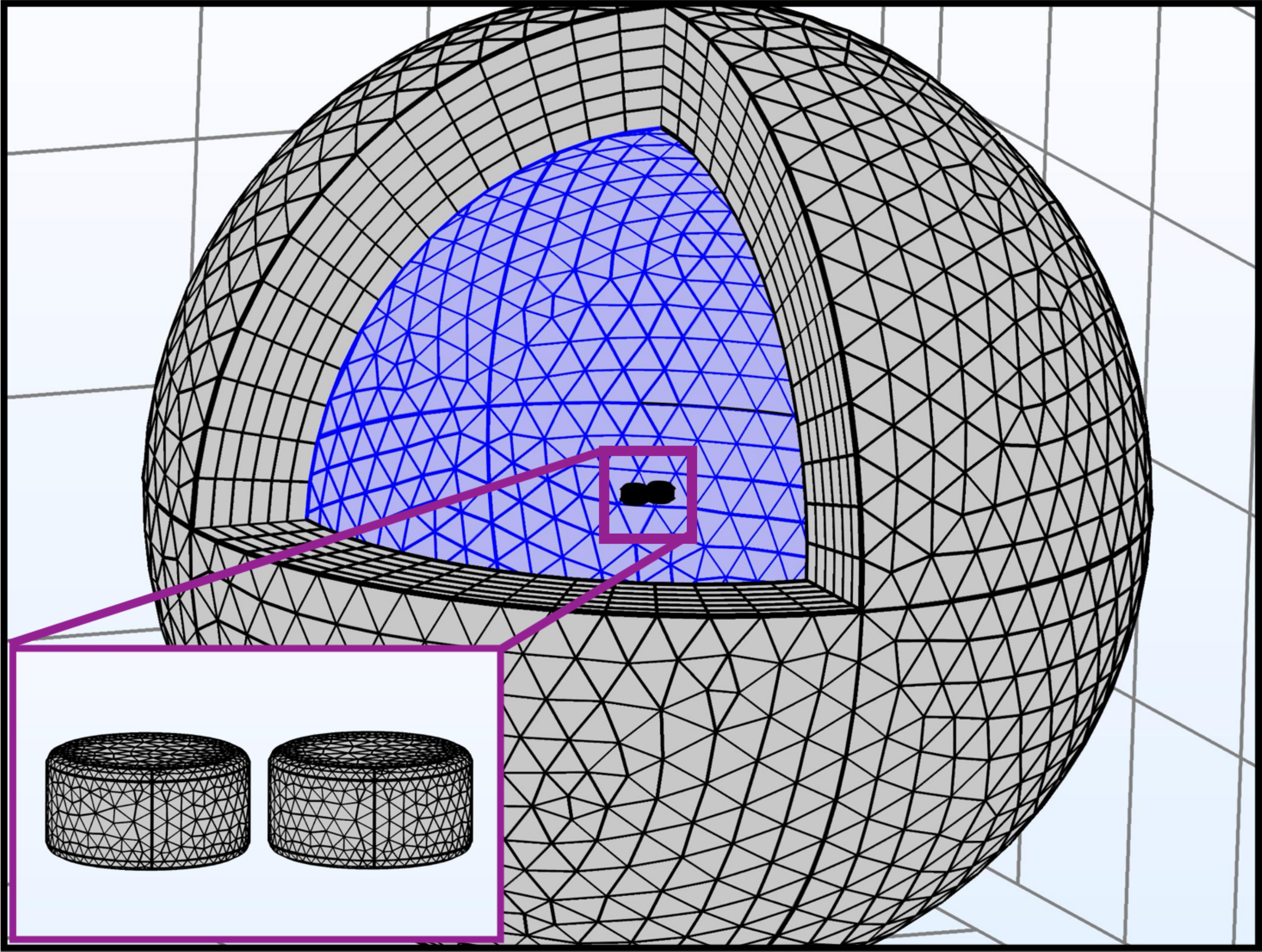}
    \caption{Modelling domain for the gold nanoparticle dimer. The modeling environment consists of air (indicated in blue), which i truncated by PML absorbing boundaries.}
    \label{fig:methods}
\end{figure}

\section{METHODS}

The numerical study was performed using a commercial electromagnetism solver tool COMSOL Multiphysics (Version 5.5) \cite{comsol5-5}. We used the \textit{wave optics} module to solve Maxwell equations in our simulation domain for the wavelength range 500--1100~nm. The dielectric function for gold was taken from Johnson and Christy \cite{J&C}. The nanoparticle dimer was placed in the middle of a hollow multi-layered core-shell structure (Fig.~\ref{fig:methods}), which consists of the modelling region. Air were used as the background index and perfectly matched layer absorbing boundary conditions were applied to prevent reflections at the simulation boundaries. Optical properties of the gold dimer were calculated by performing electro-magnetism (EM) calculations in the scattered field formulation, which has yielded reliable results for optical properties of plasmonic nanoparticles \cite{Crompton}. The background EM field was taken into account in absence of the scatterer (gold dimer), and the scattered field was then solved for the presence of the scatterer in the modelling domain. The total field consisted of the background and the scattered field together. The scattering cross-sections of the nanoparticle dimer were calculated by integrating a sphere over it and applying scattering boundary conditions. The reliability of our model was verified through by comparison with Mie theory for the visible wavelength range (see Supplementary Note 1). The corners and edges of nanodisc and nanocube dimers were rounded by a radius of 5~nm to counter the EM field effect related artefacts, which can result due to sharp edges or corners of the nanoparticle. A blue-shift in the scattering spectra  was observed as a consequence of rounding the edges of the nanocube dimer (see Supplementary Note 6), which has been found in agreement with existing studies detailing the effect of rounding on the scattering properties of nanocubes \cite{Raziman_2}.    

\begin{figure}
    \centering
    \includegraphics[width=0.45\textwidth]{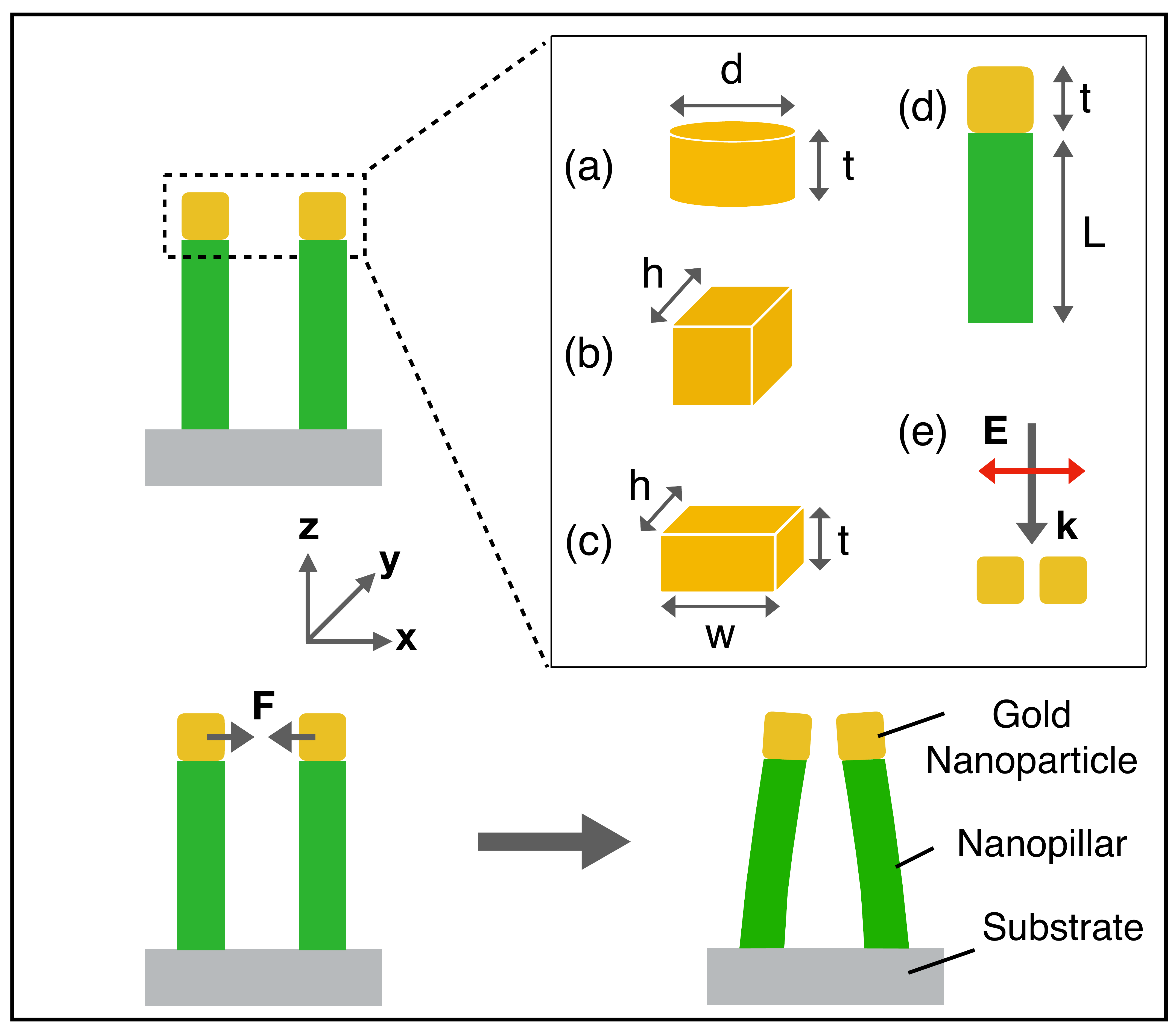}
    \caption{Schematics of plasmomechanical nanopillar antenna system. \textbf{a} Nanodisc antenna of diameter $d$ and thickess $t$. \textbf{b} Nanocube antenna of edge-length $h$. \textbf{c} Nanobar antenna of edge-length $h$, thickness $t$ and length $w$. \textbf{d} The entire pillar structure comprised of nanopillar of length $L$ and gold nanoparticle of thickness $t$. \textbf{e} Excitation conditions of the nanoantenna dimer system with electric field $E$ and wave vector $k$. The bending of the pillars has been exaggerated here for better illustration.}
    \label{fig:schematics}
\end{figure}

\section{RESULTS \& DISCUSSION}
For calculation of optical forces in plasmomechanical nanopillar system, light was perpendicularly incident on the gold nanoparticle dimer system with electric field aligned along the inter-particle axis of the dimer (see schematic in Fig. \ref{fig:schematics}). Optical forces were first calculated for a gold nanodisc dimer as it represents the simplest geometry for a  plasmomechanical system (i.e. plasmonic nanoparticle on top of a cylindrical non-metallic nanostructure). The scattering properties were studied in gold nanodisc dimer of diameter $d$ for changes in the optical properties due to plasmonic coupling at varying inter-particle distances between the nanodiscs. Scattering efficiencies of nanodisc dimers of diameter 100~nm, 150~nm and 200~nm were calculated under plane wave excitation conditions for inter-particle distances ranging between 10--50~nm. The plasmon resonance peak associated with the scattering spectra red-shifted as the distance between the two nanodiscs decreased for all radii (see Supplementary Note 2). The \textit{Maxwell stress tensor} method was used for calculation of the optical forces on the gold dimers. This method has been implemented multiple times for its reliability for calculation of optical forces in plasmonic nanostructures \cite{Oliver,Ploschner}. In contrast to the conventional approach of enclosing a box over the nanoparticle's surface and calculating the forces on it, the force tensor components were integrated directly over the gold nanodisc surface. The surface integration allows capturing of arbitrariness related to the shape of the nanostructure, which can greatly influence the magnitude of the optical forces arising from it. This approach has previously been reported to estimate the optical forces to a greater accuracy on realistic nanoparticles and molecules \cite{Ji}. Optical forces were studied between disc dimers for three different diameters 100~nm, 150~nm and 200~nm with a thickness $t$ of 50~nm.  The forces were calculated along the inter-particle axis between the discs, which makes the x-component of the stress tensor $F_x$ relevant  for our calculations. We kept a minimum distance of 10~nm between the nanostructures in our studies as multi-polar order plasmon coupling prevails at shorter distances \cite{Nasreen, Miljkvic}, which could interfere with the nature of the attractive optical forces. In order to establish effective plasmonic coupling between the nanoparticles, the polarisation of the incident light was aligned along the inter-particle axis of the dimer. The excitation wavelengths were chosen for each case corresponding to their plasmon resonances, as local field enhancement maximises at resonant excitation.

\begin{figure}
\centering
\includegraphics[width=0.485\textwidth]{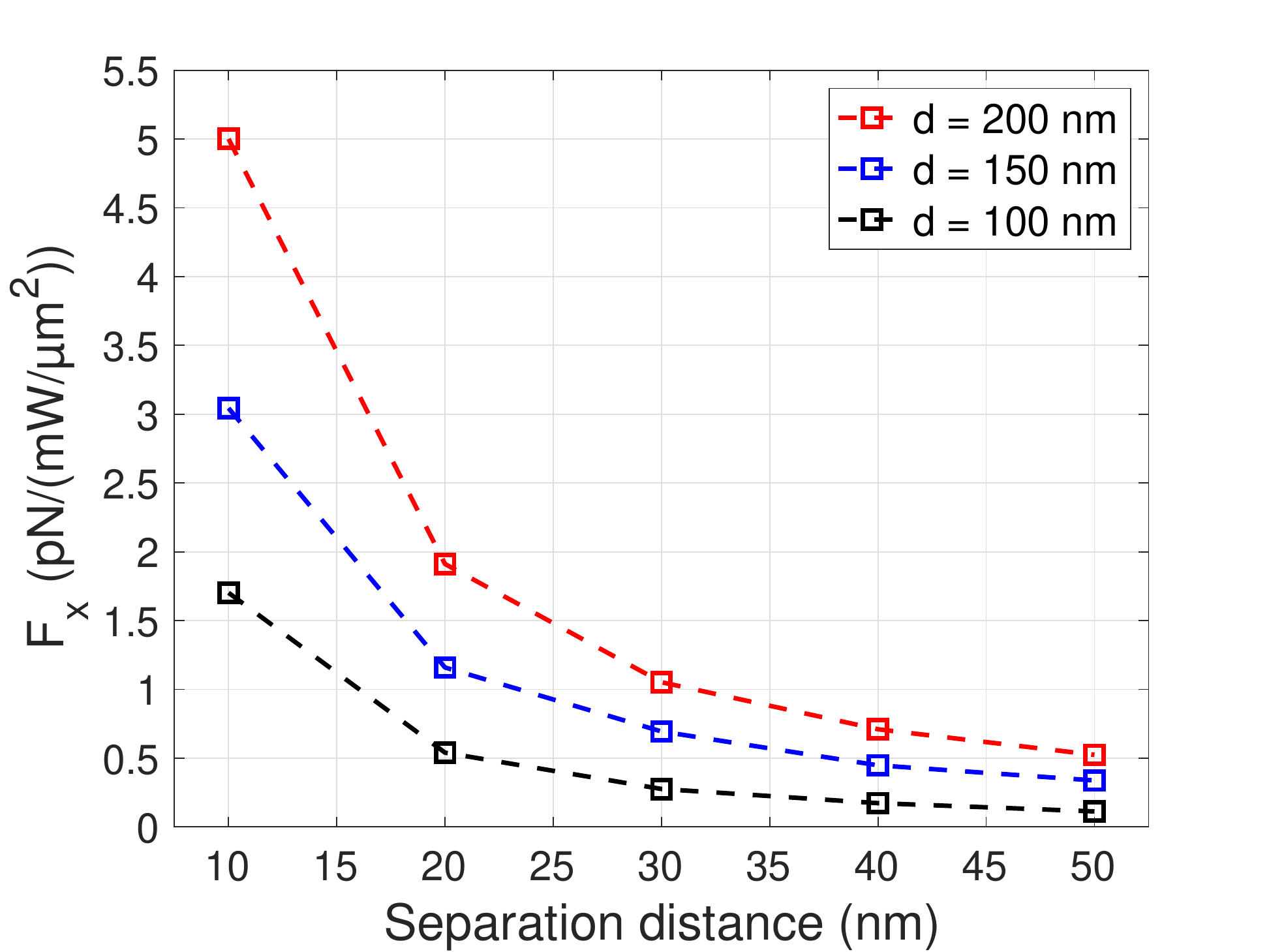}
\caption{Optical forces in gold nanodisc dimers of diameter 100~nm, 150~nm and 200~nm, with 10--50~nm inter-particle distance between the discs. Nanodisc dimers were excited under plane wave conditions with electric-field strength of 1~V/m. Each nanodisc dimer was excited at its respective plasmon resonance, deduced from its scattering spectra (given in Supplementary Note 2).}
\label{fig:discs_Fx}
\end{figure}

As shown in (Fig.~\ref{fig:discs_Fx}), the optical forces grow in magnitude as the distance between the two nanodiscs decreases. Moreover, the influence of radius of the nanodiscs can be seen on the optical forces, as the optical forces grow further in magnitude with increasing radius of the nanodiscs for all separation distances. The largest attractive force of 5~pN/(mW/$\mu$m$^2$) was calculated for the nanodisc of radius 100~nm at 10~nm separation distance. The increase in optical forces with decreasing inter-particle distances follows as a consequence of greater plasmonic coupling between the discs, which has been previously observed for bipyramidal gold nano-dimers \cite{Nome}. We calculated the local field enhancement of the electric field ($E=E_{np}/E_0$), where $E_{np}$ is enhancement of the nanoparticle and $E_0$ is the incident field enhancement, in 100~nm, 150~nm and 200~nm diameter nanodiscs respectively, to observe plasmonic coupling between the nanodiscs. A stronger plasmonic coupling can be seen to emerge at smaller gap distances between nanodiscs for all radii, which increases the local field enhancement (see Supplementary Note 3). To understand the nature of the optical forces exhibited by nanodisc dimers, we studied charge distribution on their surfaces. On exciting nanodiscs at their plasmon resonance conditions, opposite charged centers form on both nanodiscs on their upper and lower edges. The increase in the diameter of the nanodiscs leads to an increase in their polarisability, with a polarisation density of $4 \times 10^{-12}$~C/m$^2$ and $20 \times 10^{-12}$~C/m$^2$ in 100~nm and 200~nm diameter discs respectively, as observed from the polarisation density strength of the nanodiscs (see Supplementary Note 4). The distribution of charges on the nanodisc seems to be unaltered from the change in diameter. The increase in the dipolar strength with increasing diameter of the nano-dimer causes a stronger electrostatic force between nanodiscs of larger radii, resulting in a larger attractive force between them.

The plane wave approximation condition often complies with a coherent source with an illumination area large enough to substantially exceeds the geometric scattering cross-sections of the scatterer under exposure. In real world scenarios however, we often come across light sources with a Gaussian profile and a limited spot size. To draw relevance of our studies with excitation source conditions used in experimental set-ups, we calculated optical forces in the nanodiscs using Gaussian beam excitation conditions. We observed striking similarities in the inter-particle gap dependent trends and the overall magnitude of optical forces obtained by Gaussian beam excitation with that of optical forces calculated using plane-wave excitation conditions (see Supplementary Note 5). An attractive force of 4.57~pN/(mW/$\mu$m$^2$) was calculated in 100~nm radius nanodisc dimer under Gaussian beam excitation as compared to 5~pN/(mW/$\mu$m$^2$) under plane wave excitation. The observations provide a reasonable validation for exploring plasmonic enhanced optical forces under plane wave excitation conditions.   

The next set of numerical studies on internal optical forces was performed on gold nanocube dimers as they display great potential for localisation and enhancement of the local electric field. Nanocube dimers of varying sizes and configuration have previously been explored for their ability to strongly confine the electromagnetic field around their corners and in the gap region between them \cite{Bordley,Hooshmand}. Moreover, the localisation of the electric field enhancement in nanocubes can be externally tuned by varying the wavelength of the excitation source \cite{Mostafa}. The nanocube's edge length $h$ and inter-particle distance play an important role in realising the wavelength related tunability. The onset of wavelength based tunable effect was observed for nanocubes with edge-length exceeding 75~nm with inter-particle distance of 10~nm for a nanocube dimer arranged in a side-side configuration. We compared near-field enhancements in the nanocube dimers of edge length 100~nm and 150~nm with 10~nm inter-particle gaps. The maximum electric-field enhancement was achieved at resonant excitation wavelengths for both edge-lengths, with strong localisation of the electric-field present in the gap region between the nanocubes. On red-shifting the source wavelength, the hot spots moved in position towards the outer edges of the cube-dimer (see Supplementary Note 7).    

\begin{figure}
     \centering
     \includegraphics[width=0.485\textwidth]{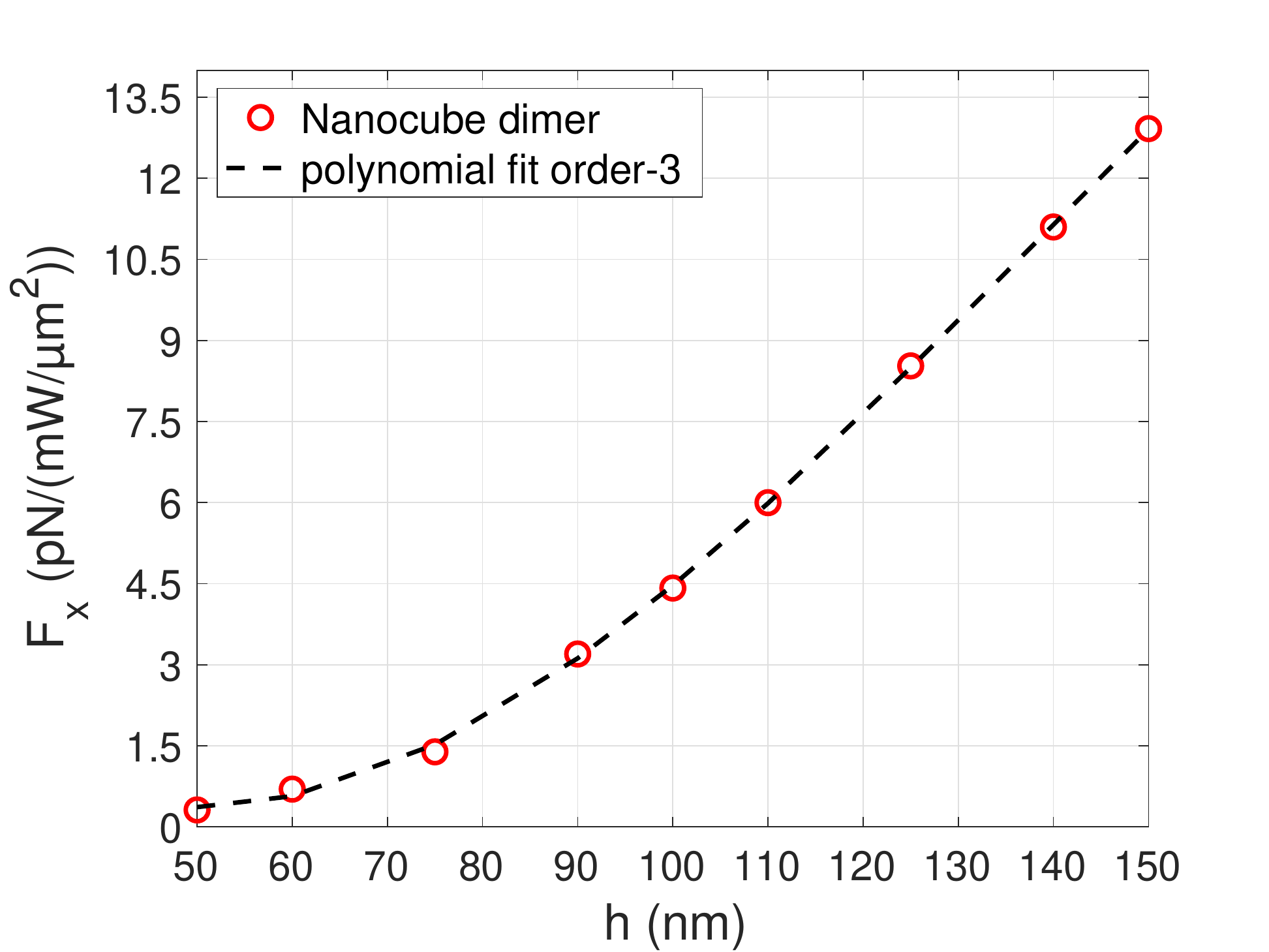}
     \caption{Optical forces in gold nanocube dimers of edge-length 50--150~nm at 10~nm inter-particle gap distance. The calculations were performed under plane wave excitation conditions using a field strength of 1~V/m. All cases of nanocube dimers were excited at their respective plasmon resonances.}
     \label{fig:cubes_Fx}
\end{figure}   

To understand the effect of particle diameter on optical forces in nanocube dimers, we calculated optical forces in gold nanocube dimers by varying the edge-length, while keeping a constant separation distance of 10~nm. The excitation wavelength for each case was chosen corresponding to the plasmon resonance of the nanocube dimer. With increasing edge-length the optical forces between the nanocube dimers grow stronger in magnitude. An attractive force of 0.32~pN/(mW/$\mu$m$^2$) was present between nanocubes of 50~nm edge length, which further rises on increasing the edge-length to 150~nm as seen in (Fig.~\ref{fig:cubes_Fx}). The optical forces realised in nanocube dimers are stronger in magnitude compared to nanodiscs. For nanodisc and nanocube dimer of same diameter 150~nm and 10~nm gap distance, attractive force of 3~pN/(mW/$\mu$m$^2$) in nanodiscs and 13~pN/(mW/$\mu$ m$^2$) in nanocubes was calculated respectively. Interestingly, optical forces in nanocube dimers were seen to exhibit a wavelength specific behavior unlike nanodiscs as shown in Supplementary Note 8. The change in the optical forces vs wavelength (here now on referred to as  spectral response of optical forces), in a nanocube dimer of 50~nm edge-length exhibits a broad peak which follows its scattering spectra. On increasing the edge-length of the nanocube dimer to 125~nm, the spectra of optical forces narrows down substantially, while the scattering spectra further broadens and exhibits additional peaks, as expected due to field retardation effects experienced for increase in the size of the nanoparticle. On the other hand the spectral response of the optical forces in nanodisc dimers follow the scattering spectra closely and does not substantially deviate from it, even for increasing radii of the nanodiscs. The distribution of electric charges on the nanostructure's surface could potentially be the prominent reason for such behaviour, which in case of a nanocube dimer is altered with variation in the edge-length. The increase in nanocube's edge-length leads to the asymmetric distribution of charge over the nanocube's surface, which causes narrowing of the spectral range of the attractive forces. To support our explanation, we studied the polarisation density for a nanocube dimer with an edge-length of 125~nm and 10~nm inter-particle separation to observe the charge distribution over it. The degree of polarisation for the dimer was seen to vary substantially within a small wavelength range around its plasmon resonance, along with asymmetric distribution of the electric charges across the facing sides of the nanocubes in the dimer configuration (see Supplementary Note 9).

\begin{figure}
    \centering
    \includegraphics[width=0.485\textwidth]{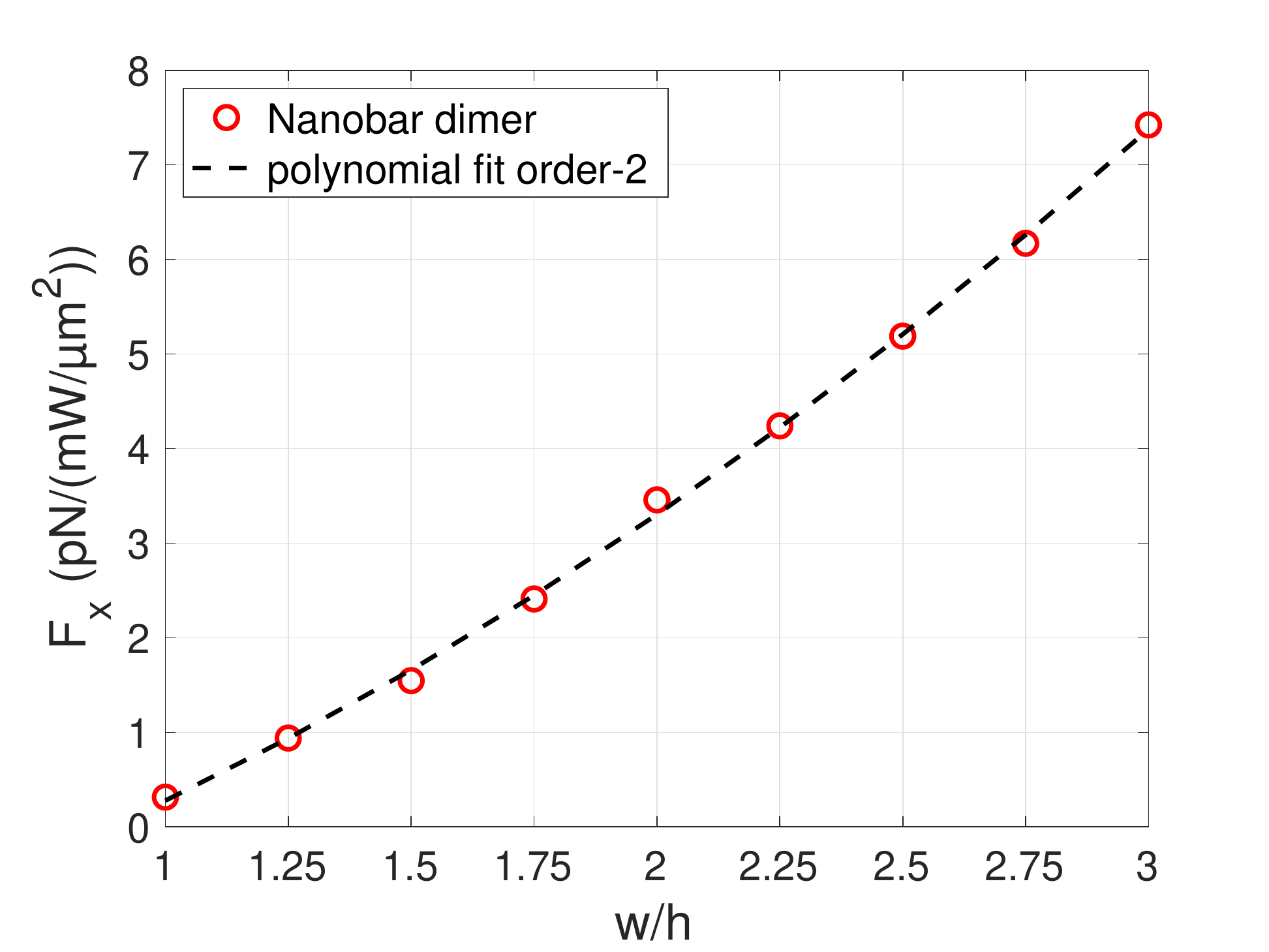}
    \caption{Optical forces in gold nanobar dimers of side length 50~nm, height 50~nm and length varied between 50--150~nm at 10~nm inter-particle gap distances. The calculations were performed under plane wave excitation conditions using a field strength of 1~V/m. All cases of nanobar dimers were excited at their respective plasmon resonances.}
    \label{bars:Fx}
\end{figure}

We extended our studies of optical forces to the case of a nanobar dimer to gain a better quantitative understanding of the relation between charge distribution on the nanostructure's surface and the spectral response of the resulting optical forces. Studying optical forces for a nanobar configuration allows the possibility to vary the length of the nanostructure while keeping the edge length and thickness constant. Choosing this geometry therefore offers insights into changes in distribution of charge on the nanobar dimer by varying its aspect ratio. The side-length $h$ and thickness $t$ of the nanobar dimer was fixed to 50~nm at 10~nm inter-particle separation, while the length $w$ was varied from 50~nm to 150~nm. As seen in (Fig.~\ref{bars:Fx}), the optical forces increase linearly with an increase in the aspect ratio, with an attractive force of 7.5~pN/(mW/$\mu$m$^2$) experienced by the nanobars of length 150~nm. This force is greater in magnitude in comparison to the strongest force observed in nanodisc 5~pN/(mW/$\mu$m$^2$). The stronger optical force in nanobar dimer stems from its higher dipolar strength of $26 \times 10^{-12}$~C/m$^2$, as compared to $20 \times 10^{-12}$~C/m$^2$ in nanodiscs, resulting in a stronger electrostatic interaction between the nanobars (see Supplementary Note 10). The spectral response of the optical forces in a nanobar dimer exhibits a behaviour similar to what we observed in nanodisc dimers; it follows the scattering spectra closely, even for increasing aspect ratio of the nanobars (see Supplementary Note 8). The charges calculated  across the nanobar dimer of highest aspect ratio 3, exhibit a dipolar like charge distribution with symmetric charge centers acquired by both nanobars (see Supplementary Info 10). Since varying the aspect ratio does not alter the side-length of the nanobars in our case, the symmetry of the charge distribution seems to be unaffected by changes made to the length. This observation reinforces our explanation that the narrow spectral response of optical forces follows as a consequence of the asymmetric distribution of the charges present on the nanostructure's surface, which is uniquely displayed by nanocube dimers.

The irradiance from the excitation source plays an important role in enhancing light-matter interactions in plasmonic nanostructures and can strongly influence the optical forces resulting from it, given the emitted power levels do not exceed the damage threshold of the structures under illumination. Metal nanostructures, in-particular gold, can resist changes to its optical properties upon radiant exposure upto a few hundred milli-watts \cite{Summers}, which makes it a promising candidate for studying effects of increasing power on the optical forces. Choosing an optimal geometry is equally important to fully explore the potential for boosting the optical forces in nanostructures. We compared the enhancement of optical forces in a gold nanodisc dimer of 100~nm diameter with a nanocube dimer of 100~nm edge-length, since the regions of electric field localisation differs in both cases due to geometry. The inter-particle distance for the dimer was kept 20~nm and  the irradiance was controlled by using a Gaussian beam source. Usually the spot-size of Gaussian beams used in numerical studies are limited to micron sizes for visible wavelength ranges. Narrowing the beam spot to the order of the excitation wavelength offers the possibility for improving optical forces as a consequence of increase in the electromagnetic energy density. We resolved this constraint by employing the non-paraxial beam approximation condition for our Gaussian beam source, which has been introduced in recent version of COMSOL Multiphysics (Version 5.6), allowing 1:1 ratio of the beam spot to the excitation source wavelength. This formulation utilises the plane wave expansion methodology to generate a Gaussian beam with a tighter spot size (see Supplementary Note 11). The dimer nanoantennas were excited at their respective plasmon resonance conditions, which is 625~nm for the nanodisc and 670~nm for the nanobar. The deviation of the beam spot between the two cases of excitation was small enough to not externally influence the magnitude of the optical forces.
\begin{figure}
    \centering
    \includegraphics[width=0.485\textwidth]{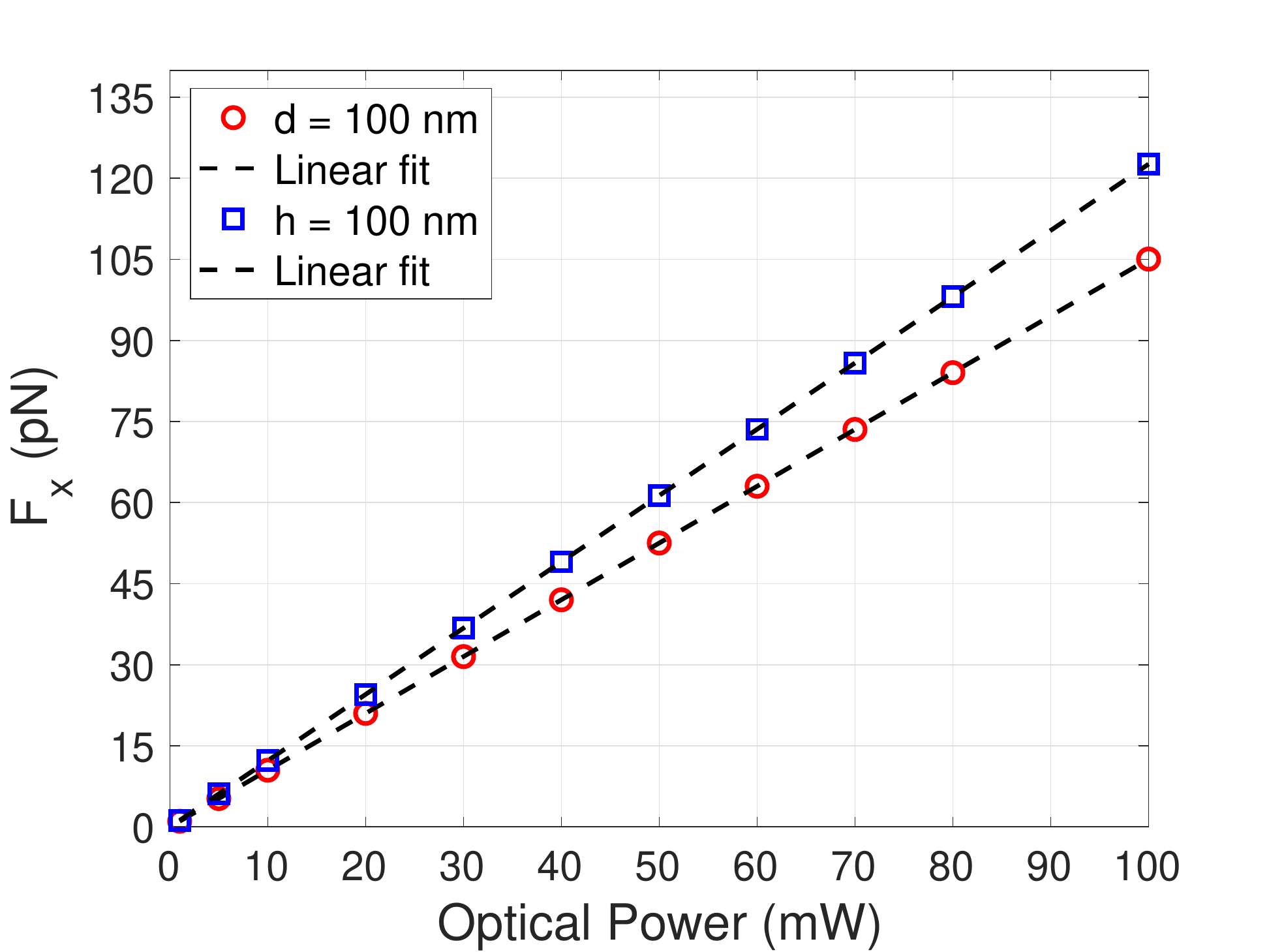}
    \caption{Optical forces in nanodisc dimer of radius 100~nm and nanocube dimer of edge-length 100~nm at 20~nm inter-particle distance. The calculations were performed under Gaussian beam excitation conditions in the non-paraxial approximation. nanodiscs were excited at 625~nm and nanobars at 670~nm. The spot size of the Gaussian beam was kept equivalent to the excitation wavelength for both cases respectively.}
    \label{fig:power_Fx}
\end{figure}

The relation of the attractive optical forces with power levels of the excitation source for a nanodisc and nanocube dimer can be seen in  (Fig.~\ref{fig:power_Fx}). It can be seen that attractive forces between the nanostructures increases linearly with the optical power, and this behaviour is exhibited by both nanostructure geometries. It is evident that attractive forces between the nanocube dimer grow stronger with increasing power levels compared to the nanodiscs. However, at low optical power excitation upto 10~mW, there is a similar increase in the optical forces for both geometries. The local field enhancement in a nanocube dimer is stronger than in nanodiscs, which consequently leads to exertion of a stronger optical force in them. The attractive forces can reach upto 120~pN in a gold nanocube dimer and 100~pN in case of a gold nanodisc dimer for 100~mW power of the excitation source. Such magnitudes of attractive optical forces show great potential for all-optical modulation such as actuation and transduction of plasmomechanical systems at the nanoscale.
 
The transduction of motion in nanopillars is decisive for their applications. Optical force acting on a plasmomechanical nanopillar can be approximated as a point load since the dimensions of a gold nanoparticle are quite smaller compared to the nanopillar itself. In the first bending mode of nanopillars, the maximum displacement amplitude lies at the free end of the pillar resonator \cite{Schmid}, which goes well with placement of the gold nanoantenna on top of the nanopillar for producing maximum deflection. The stiffness of the nano-pillar is governed by its geometric moment of inertia, which depends on the pillar geometry. Since the motion of the nanopillars is along x-axis due to optical forces, the area moment of inertia for bending in $x$-direction is expressed as $I_y = \iint z^2$~dA. The deflection of pillars $x = F/k $ can be estimated from the ratio of the force applied $F$ to the stiffness of the pillars $k = 3 E I_y/L^3$, with the geometric moment of inertia $I_y = \pi d^4/64$ and $I_y = h^4/12$ of a disc and cube, respectively.
 
\begin{table}
\caption{\label{tab:table1}Estimated deflection in silicon nanopillars with gold nanodimer on top and 10~nm inter-particle distance. Plane wave excitation conditions were used with 1 V/m electric-field strength.   %
}
\begin{ruledtabular}
\begin{tabular}{cc|cc}
\textrm{$d$~(nm)}&
\textrm{$x$~(nm/(mW/$\mu$m$^2$))}&
\multicolumn{1}{c}{\textrm{$h$~(nm)}}&
\textrm{$x$~(nm/(mW/$\mu$m$^2$))}\\
\colrule
100 & 0.770 & 50 & 1.344\\
150 & 0.272 & 75 & 1.171\\
200 & 0.141 & 100 & 1.178\\
- & - & 125 & 0.931\\
- & - & 150 & 0.680\\
\end{tabular}
\end{ruledtabular}
\end{table}
 
For estimation of maximum deflection in a nanopillar dimer, we compared nanodiscs and nanocube dimers of varying diameter at 10~nm inter-particle distance. Assuming silicon nanopillars with youngs modulus $E = 150$~GPa, we chose the length of pillars $L = 10~\mu m$ for our study, since a high dynamic range of 85--90 has been recently reported for pillars around this height \cite{Molina}. As shown in table \ref{tab:table1}, the magnitude of deflection in nanodisc dimers decreases with increase in the particle diameter, with a static deflection of 0.77~nm/(mW/$\mu$m$^2$) observed for 100~nm diameter pillars. A similar trend is also observed for nanocube dimers, where maximum deflection of 1.34~nm/(mW/$\mu$m$^2$) is observed for cubes with the shortest edge-length of 50~nm. Therefore, we realise stronger magnitude of displacement in nanocubes compared to nanodiscs. The observations reveal that rather than optical forces alone, the force-balance between the optical forces and the stiffness of the pillars determines the resulting deflection of the pillars. Our results indicate that designing slender pillars can produce maximum deflection of nanopillars at short inter-particle distances and favours the geometry of nanoantennas which can produce maximum local field enhancement. Our calculations reveal an improvement in deflection by 3 orders of magnitude over deflections estimated in pillar dimers actuated by electrostatic forces \cite{Kainz}. The results inform on the promising prospects of nanopillar antennas for plasmomechanical actuation and transduction, where the performance of the nano-pillars can be further optimised through improvement of their quality factors.  

\section{CONCLUSION}
In summary, we presented detailed studies on optimising optical forces in gold nano-dimers. Our findings emphasize on the essence of electric-field enhancement for boosting optical forces arising from plasmonic interactions, and guide towards the design considerations which aid in maximising them. We observe the role of polarisability of gold nanoparticles as central in governing the nature of optical forces and show the strong influence of nanoparticle geometry on it. Using a novel numerical methodology, the influence of optical power was studied on the optical forces, signifying the role of the excitation source for amplifying their magnitude. Our results highlight the potential of optical forces realised in metal nanostructures for driving and reading out the motion of nanomechanical resonators, particularly when integrated with nanopillar structures. The all-optical actuation and transduction of nano-pillar antenna arrays has immense potential for ultra-sensitive mass and force sensing, optical modulation and reconfigurability, while offering low power demands and robust integration at the nanoscale.

\section*{ACKNOWLEDGMENTS}
This work is supported by the European Research Council under the European Union Horizon 2020 research and innovation program (Grant Agreement-716087-PLASMECS).

\newpage \pagebreak \cleardoublepage

\appendix

\section*{SUPPLEMENTARY INFORMATION FOR: Design Considerations of Gold Nanoantenna Dimers for Plasmomechanical Transduction}

\subsection{COMSOL model vs Mie theory}
\begin{figure}[h]
    \centering
    \includegraphics[width=0.48\textwidth]{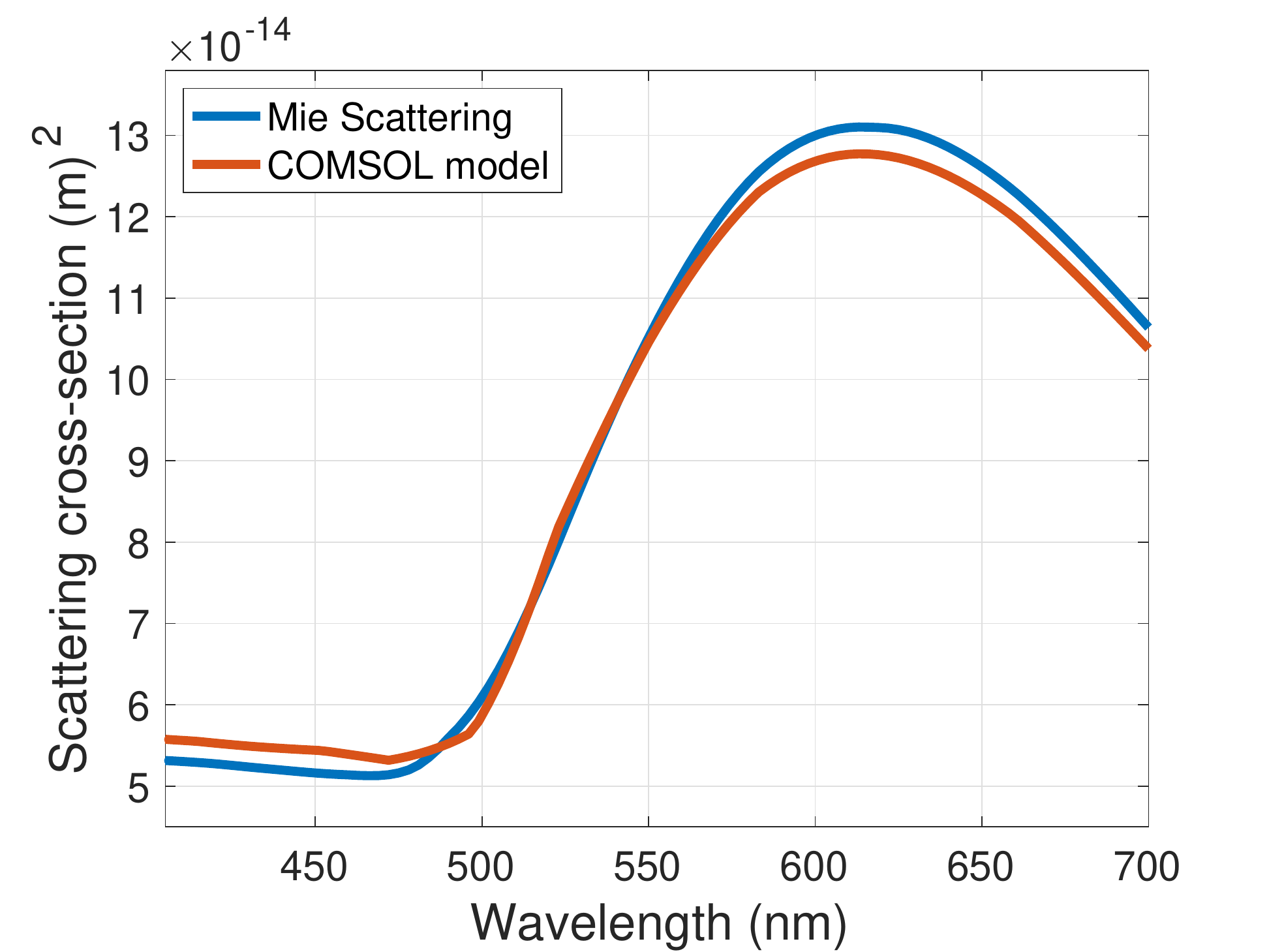}
    \caption{Comparison of scattering spectra of 100~nm diameter gold nanosphere calculated with COMSOL multiphysics with Mie scattering calculated using Mie theory.}
    \label{fig:Miescat}
\end{figure}

Scattering properties of a 100~nm diameter gold nanosphere were calculated in the visible region using dielectric function from Johnson and Christy \footnote{P. B. Johnson and R. W. Christy, Optical constants of the noble metals, Phys. Rev. B, 1972, \textbf{6}, 4370.}. In Fig. \ref{fig:Miescat} we can see a good agreement of the COMSOL model with scattering properties calculated using Mie theory.
\begin{figure}[h]
    \centering
    \includegraphics[width=0.48\textwidth]{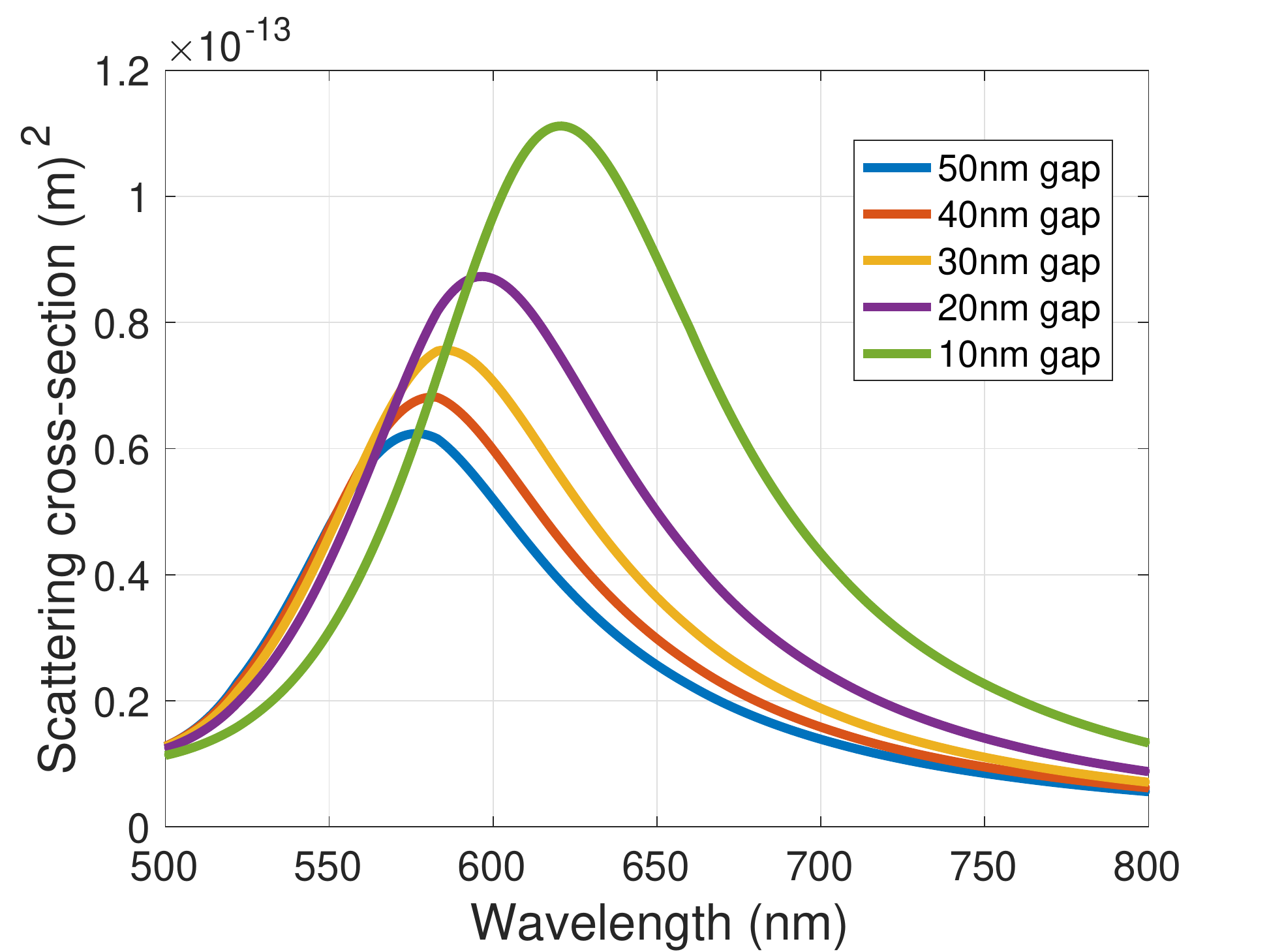}
    \caption{Scattering cross-section for a gold nanodisc dimer of radius 50~nm at varying inter-particle distances 10--50~nm under plane wave excitation.}
    \label{fig:50disc} 
\end{figure}

\begin{figure}
    \centering
    \includegraphics[width=0.48\textwidth]{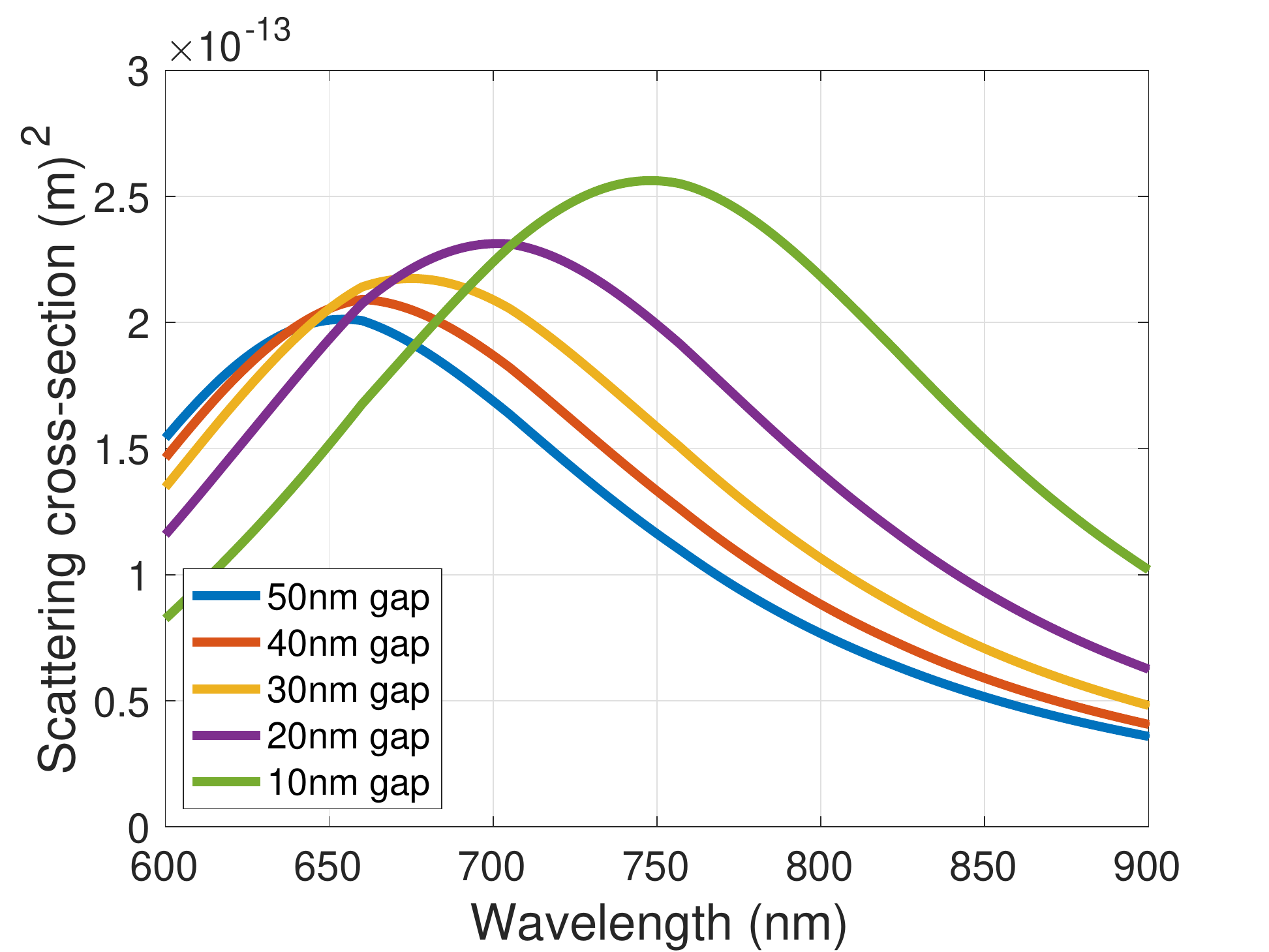}
    \caption{Scattering cross-section for a gold nanodisc dimer of radius 75~nm at varying inter-particle distances 10--50~nm under plane wave excitation.}
    \label{fig:75disc}
\end{figure}

\begin{figure}
    \centering
    \includegraphics[width=0.48\textwidth]{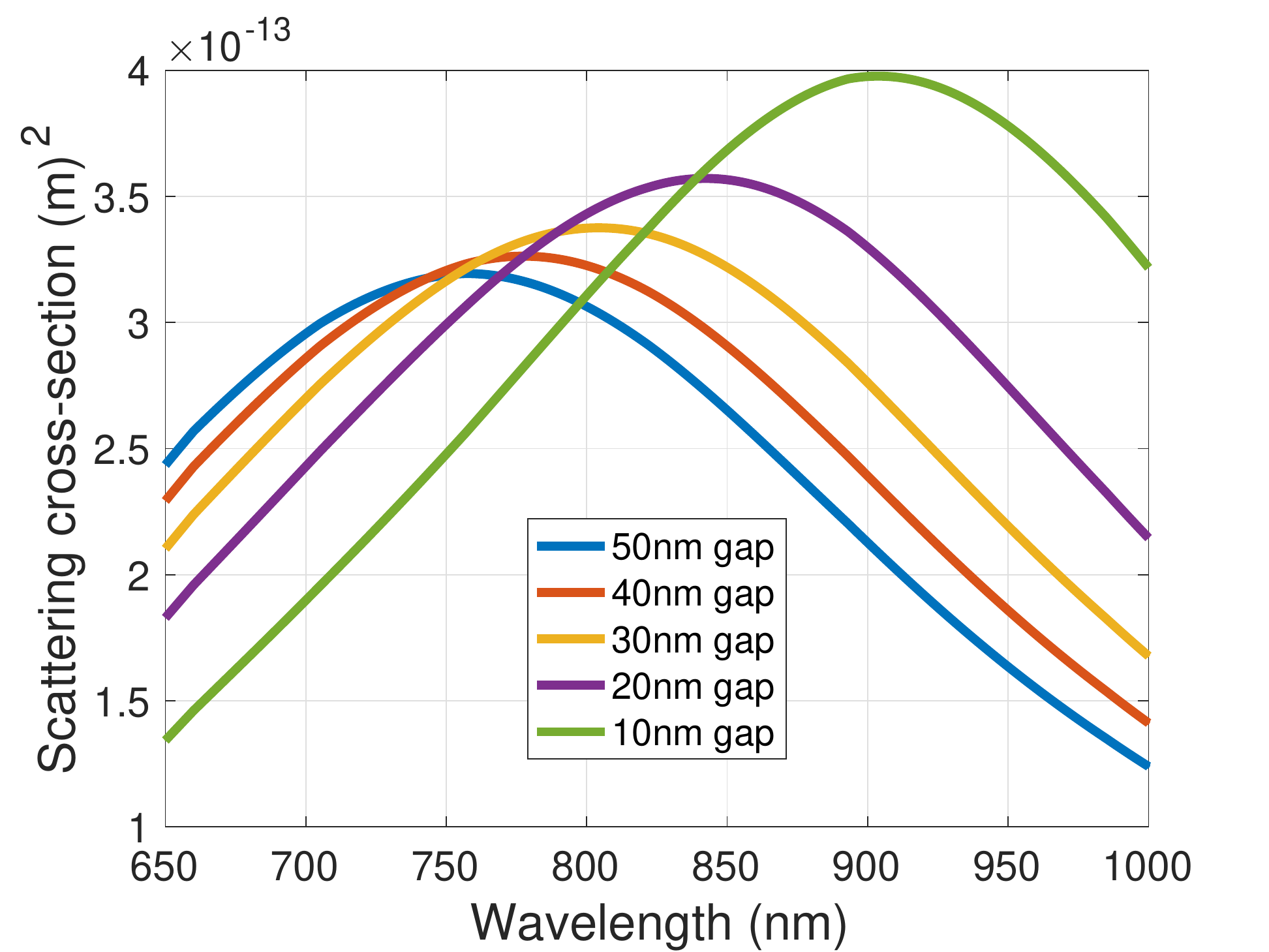}
    \caption{Scattering cross-section for a gold nanodisc dimer of radius 100~nm at varying inter-particle distances 10--50~nm under plane wave excitation.}
    \label{fig:100disc}
\end{figure}

\subsection{Scattering spectra of nanodisc dimers}
The calculated scattering spectra of nanodisc dimers of radius 50~nm, 75~nm and 100~nm are shown in Fig. \ref{fig:50disc}, \ref{fig:75disc} and \ref{fig:100disc} respectively. A red-shift in the resonance peak is observed with decreasing inter-particle distances for all dimer cases. Such behavior is displayed mainly due to increased plasmonic coupling between the nanodiscs, shifting the plasmon resonances to lower energies, accompanied with broadening of the scattering spectra. 

\begin{figure*}[t]
\begin{subfigure}[h]{0.32\textwidth}
\centering
\includegraphics[width=\textwidth]{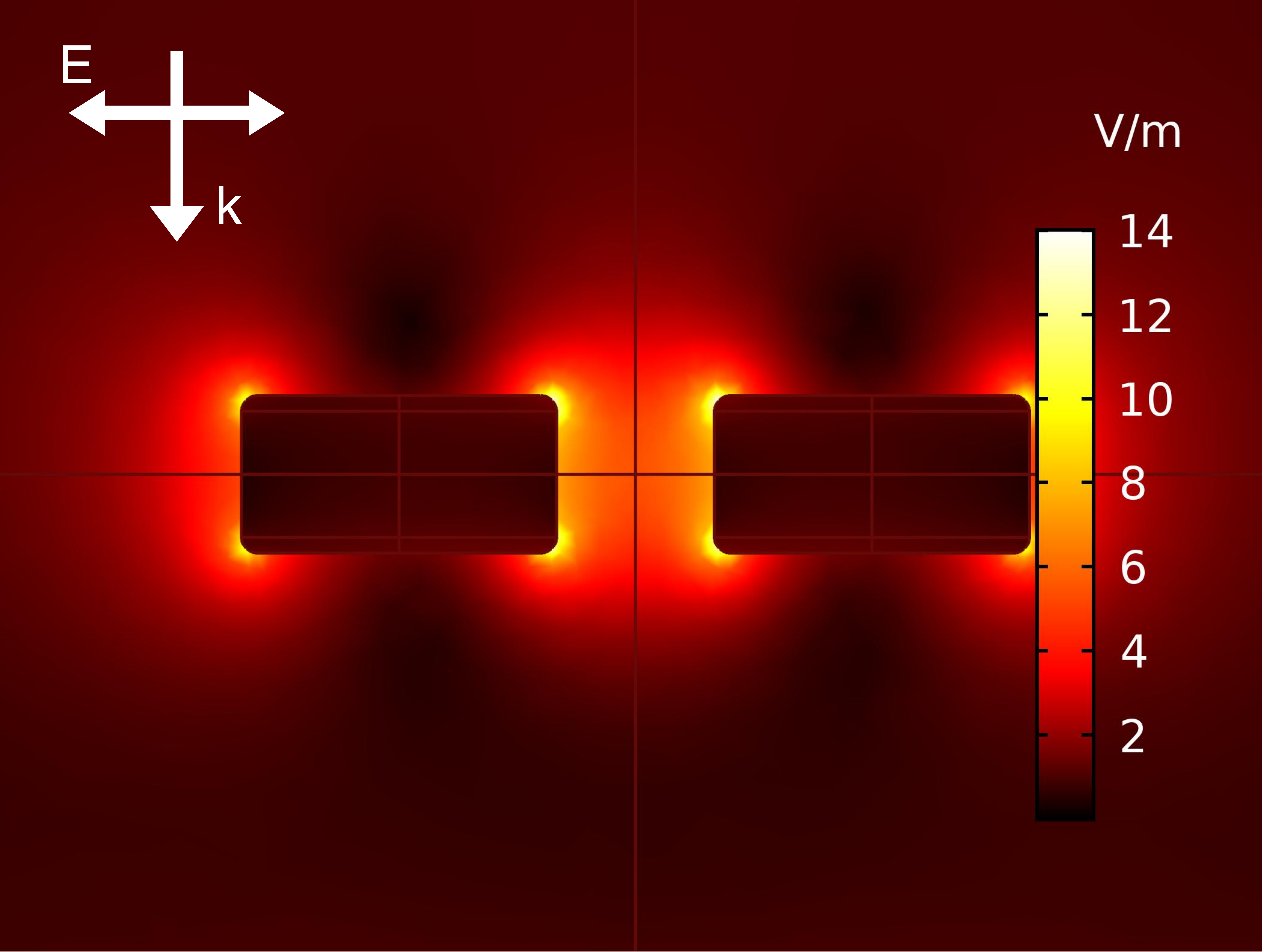}
\caption{}
\label{fig:50_50nm}
\end{subfigure}
\begin{subfigure}[h]{0.32\textwidth}
\centering
\includegraphics[width=\textwidth]{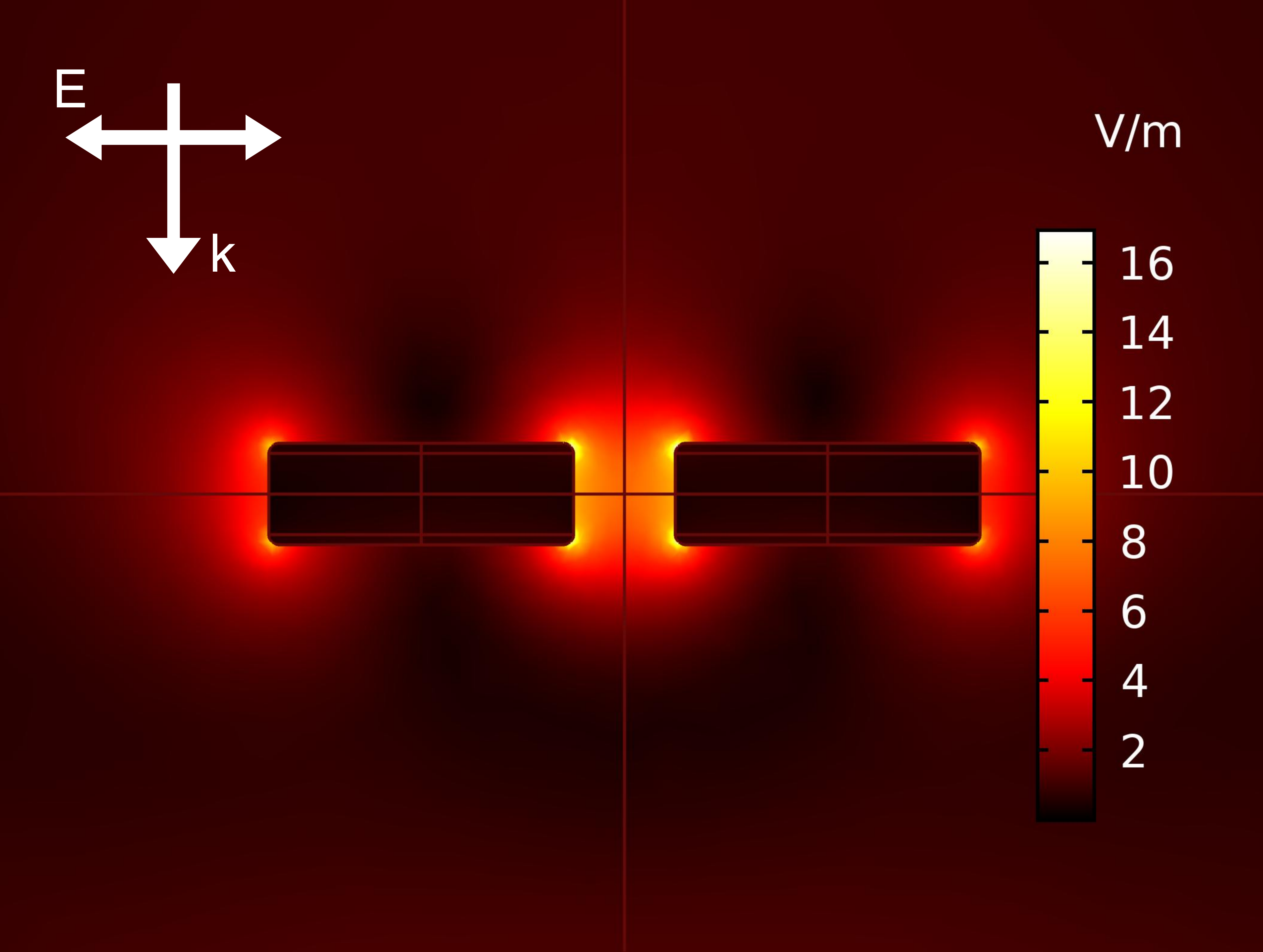}
\caption{}
\label{fig:75_50nm}
\end{subfigure}
\begin{subfigure}[h]{0.32\textwidth}
\centering
\includegraphics[width=\textwidth]{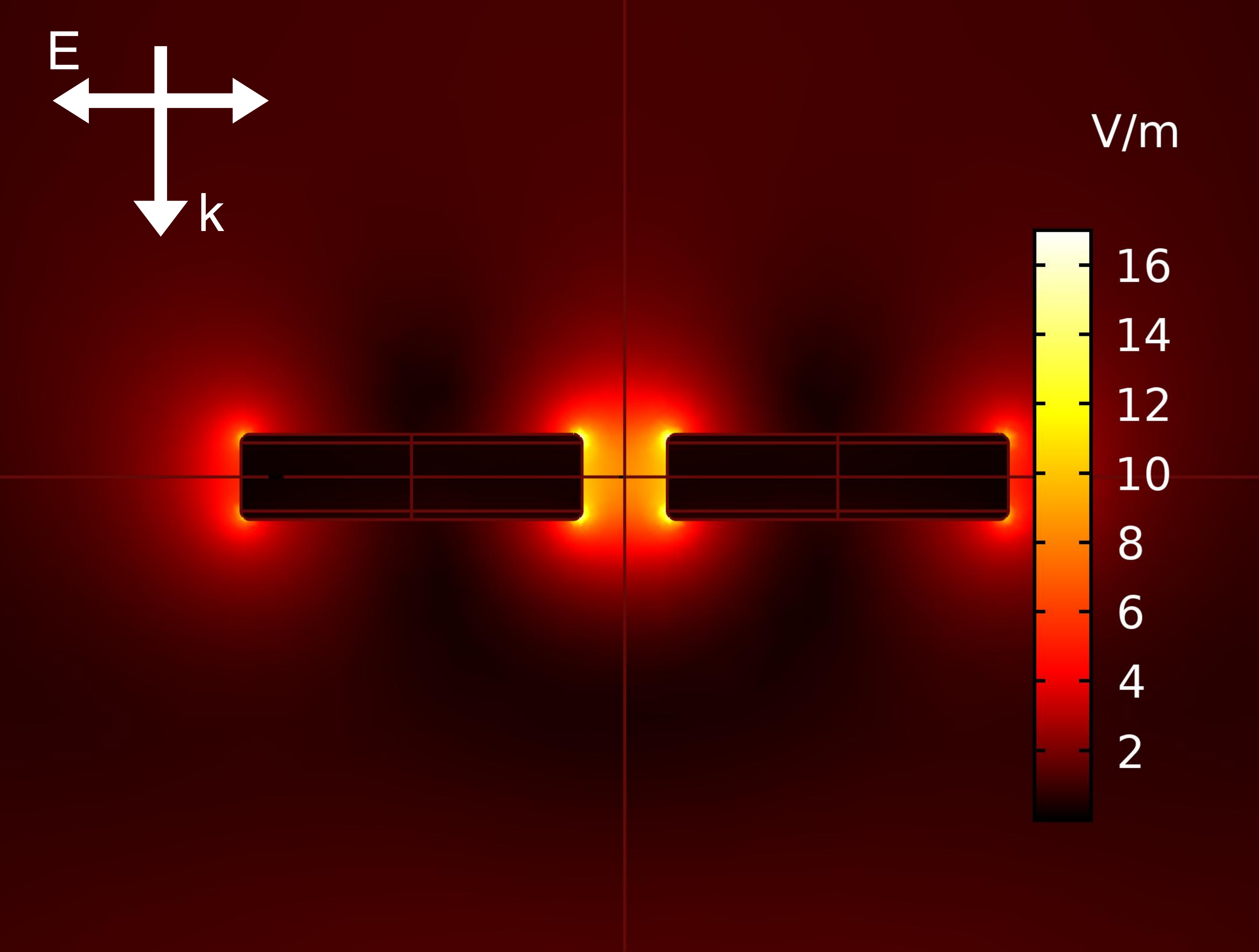}
\caption{}
\label{fig:100_50nm}
\end{subfigure}
\begin{subfigure}[h]{0.32\textwidth}
\centering
\includegraphics[width=\textwidth]{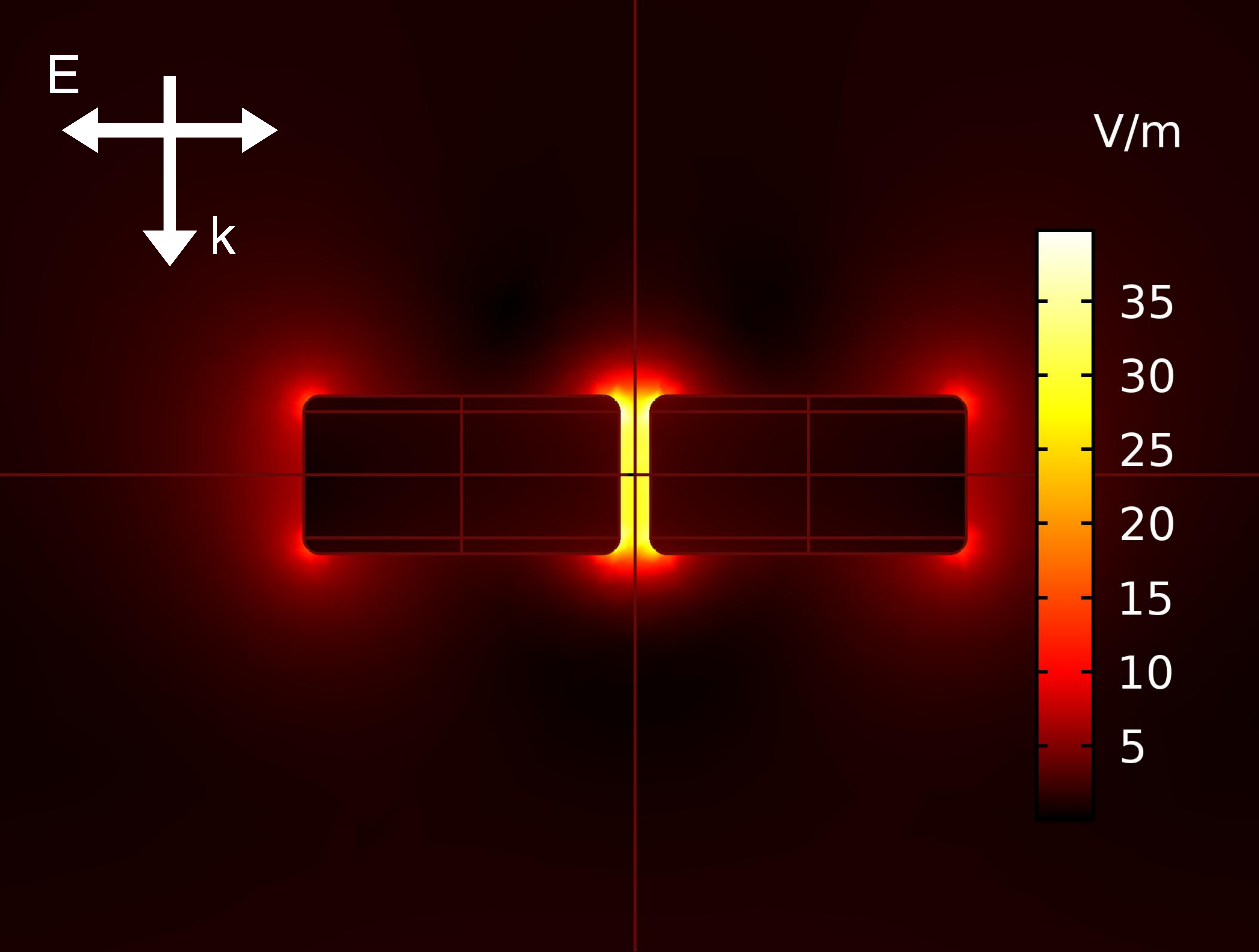}
\caption{}
\label{fig:50_10nm}
\end{subfigure}
\begin{subfigure}[h]{0.32\textwidth}
\centering
\includegraphics[width=\textwidth]{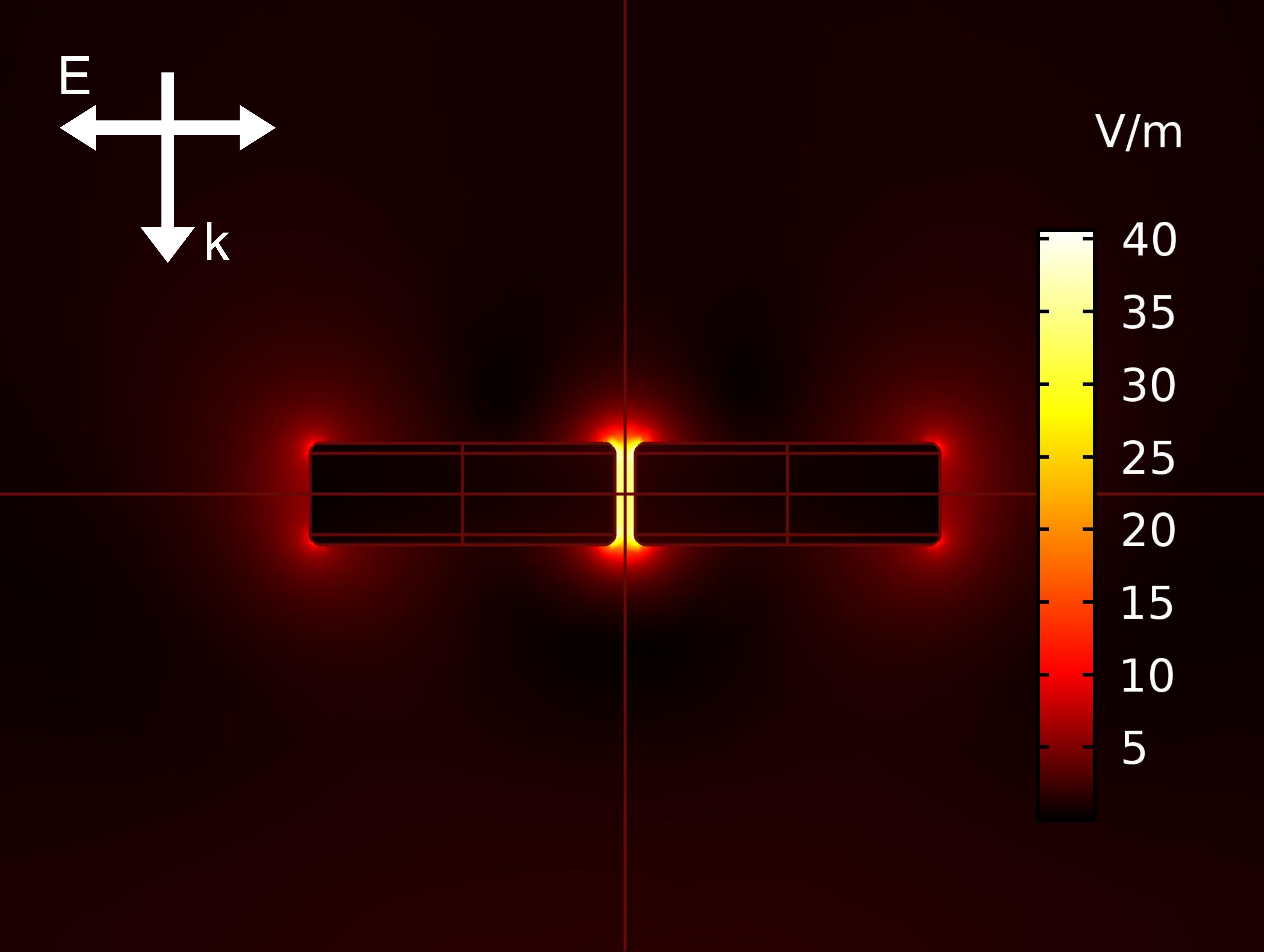}
\caption{}
\label{fig:75_10nm}
\end{subfigure}
\begin{subfigure}[h]{0.32\textwidth}
\centering
\includegraphics[width=\textwidth]{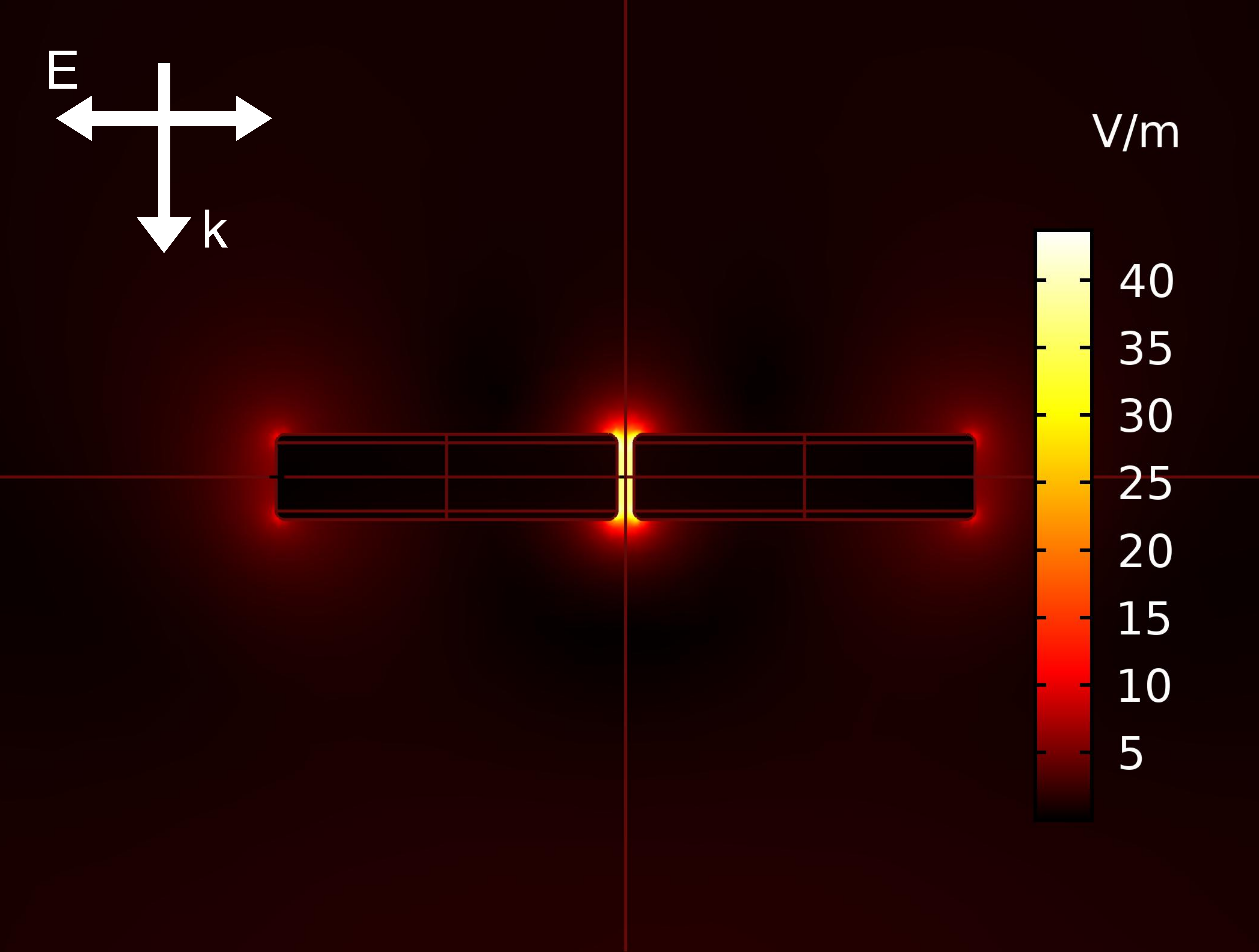}
\caption{}
\label{fig:100_10nm}
\end{subfigure}
\caption{x-z view of electric-field profile in gold nanodisc dimers under plane wave excitation normal to the plane of nanodiscs. \textbf{a} 50~nm radius nanodisc dimer with 50~nm gap at 576~nm excitation. \textbf{b} 75~nm radius nanodisc dimer with 50~nm gap at 654~nm excitation. \textbf{c} 100~nm radius nanodisc dimer with 50~nm gap at 759~nm excitation. \textbf{d} 50~nm radius nanodisc dimer with 10~nm gap at 620~nm excitation. \textbf{e} 75~nm radius nanodisc dimer with 10~nm gap at 749~nm excitation. \textbf{f} 100~nm radius nanodisc dimer with 10~nm gap at 906~nm excitation.}
\end{figure*}

\subsection{Electric field enhancement in nanodisc dimers}
Electric field enhancement was calculated for nanodiscs of radii 50~nm, 75~nm and 100~nm at gap distances of 50~nm and 10~nm respectively. Light was incident perpendicular to the plane of nanodiscs with electric field aligned along the inter-particle axis of the dimer. Fig. \ref{fig:50_50nm}, \ref{fig:75_50nm} and \ref{fig:100_50nm} show the electric field enhancement for nanodisc dimers of radius 50, 75 and 100~nm with 50~nm separation distance. Similarly, electric field enhancements at an inter-particle distance of 10~nm for discs of radius 50~nm, 75~nm and 100~nm can be seen in Fig. \ref{fig:50_10nm}, \ref{fig:75_10nm} and \ref{fig:100_10nm} respectively. The excitation wavelength for the disc dimers was determined from their scattering spectra. For each case the excitation wavelength was chosen to be at the resonance conditions. We observe a strong electric field enhancement between the discs as the distance between them decreases from 50nm to 10nm. This occurs as a consequence of stronger plasmon coupling emerging at smaller inter-paricle distances. The largest enhancement factor of 43 is observed for the 100~nm radius nanodisc dimer at a separation distance of 10~nm. The electric field enhancement for 50, 75 and 100~nm radius discs with 50~nm separation reveal a dipolar mode profile at resonance excitation conditions, with confinement regions mainly being present in the gap between the dimer and at the outer edges of the nanodisc. As the distance between nanodiscs decreases to 10~nm, the confinement in the gap region increases significantly, forming hot-spots for electric field enhancement. 

\subsection{Polarisation density maps of nanodisc dimer}
\begin{figure}
\begin{subfigure}{0.4\textwidth}
\centering
\includegraphics[width=\textwidth]{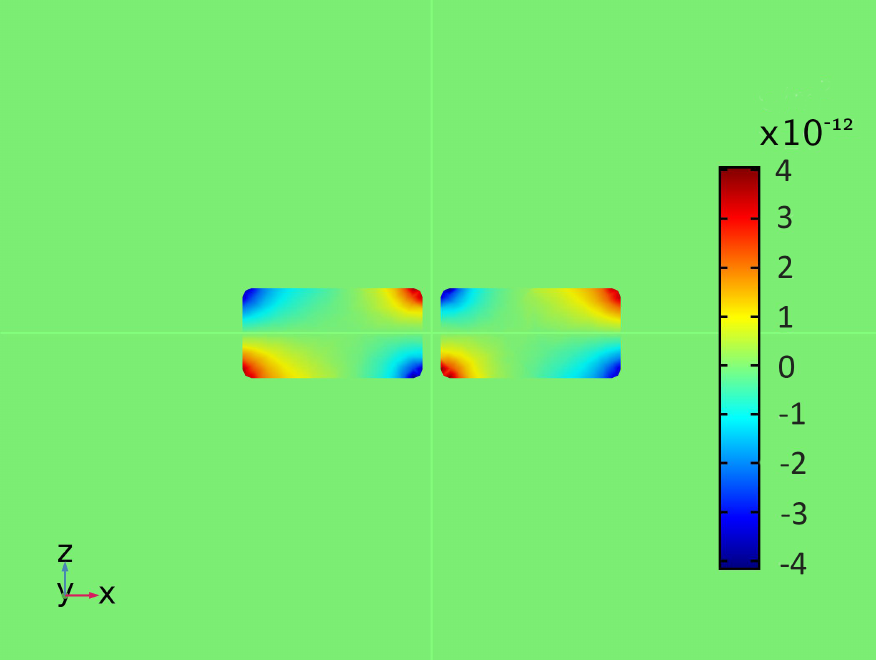}
\caption{}
\label{fig: 50pol_10d}
\end{subfigure}
\begin{subfigure}{0.4\textwidth}
\centering
\includegraphics[width=\textwidth]{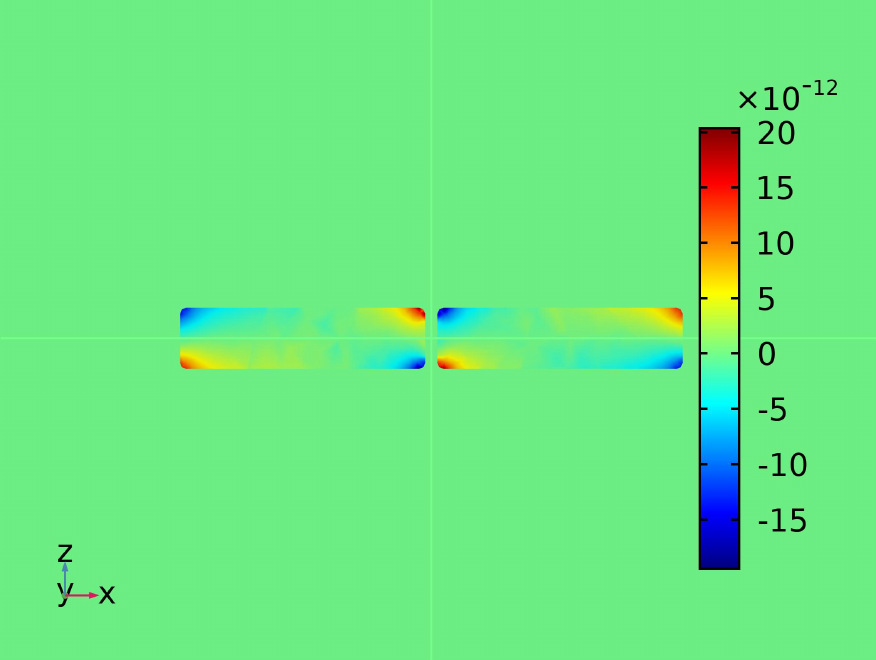}
\caption{}
\label{fig: 100pol_10d}
\end{subfigure}
\caption{Polarisation density maps of gold nanodisc dimers under plane wave excitation. \textbf{a} Polarisation density map for a 50~nm radius nanodisc dimer with 10~nm gap distance at 620~nm excitation. \textbf{b} Polarisation density map for a 100~nm radius nanodisc dimer with 10~nm gap distance at 906~nm excitation.}
\end{figure}

The polarisation maps of nanodisc dimers were obtained by calculating the charge induced on the nanodisc dimers upon optical excitation. Incident light was injected along z-axis with electric-field vector aligned along the x-axis. Fig. \ref{fig: 50pol_10d} shows the charge distribution on the 50~nm radius nanodisc dimer and Fig. \ref{fig: 100pol_10d} on 100~nm radius nanodisc dimer, with 10~nm gap between the nanodiscs. The maps reveal emergence of opposite charge centers at the top and bottom faces of the nanodiscs, which is identical for nanodiscs of both radii. The larger disc dimer of radius 100~nm attains a larger magnitude of polarisation density $20 \times 10^{-12}$~C/m$^2$ compared to $4 \times 10^{-12}$~C/m$^2$ observed in smaller nanodisc dimer of radius 50~nm. This implies that a stronger electrostatic interaction exists between the nanodiscs of larger radius, leading to a stronger attractive force between them.   

\subsection{Optical force studies in nanodisc dimers using Gaussian beam excitation}
\begin{figure}
    \centering
    \includegraphics[width=0.52\textwidth]{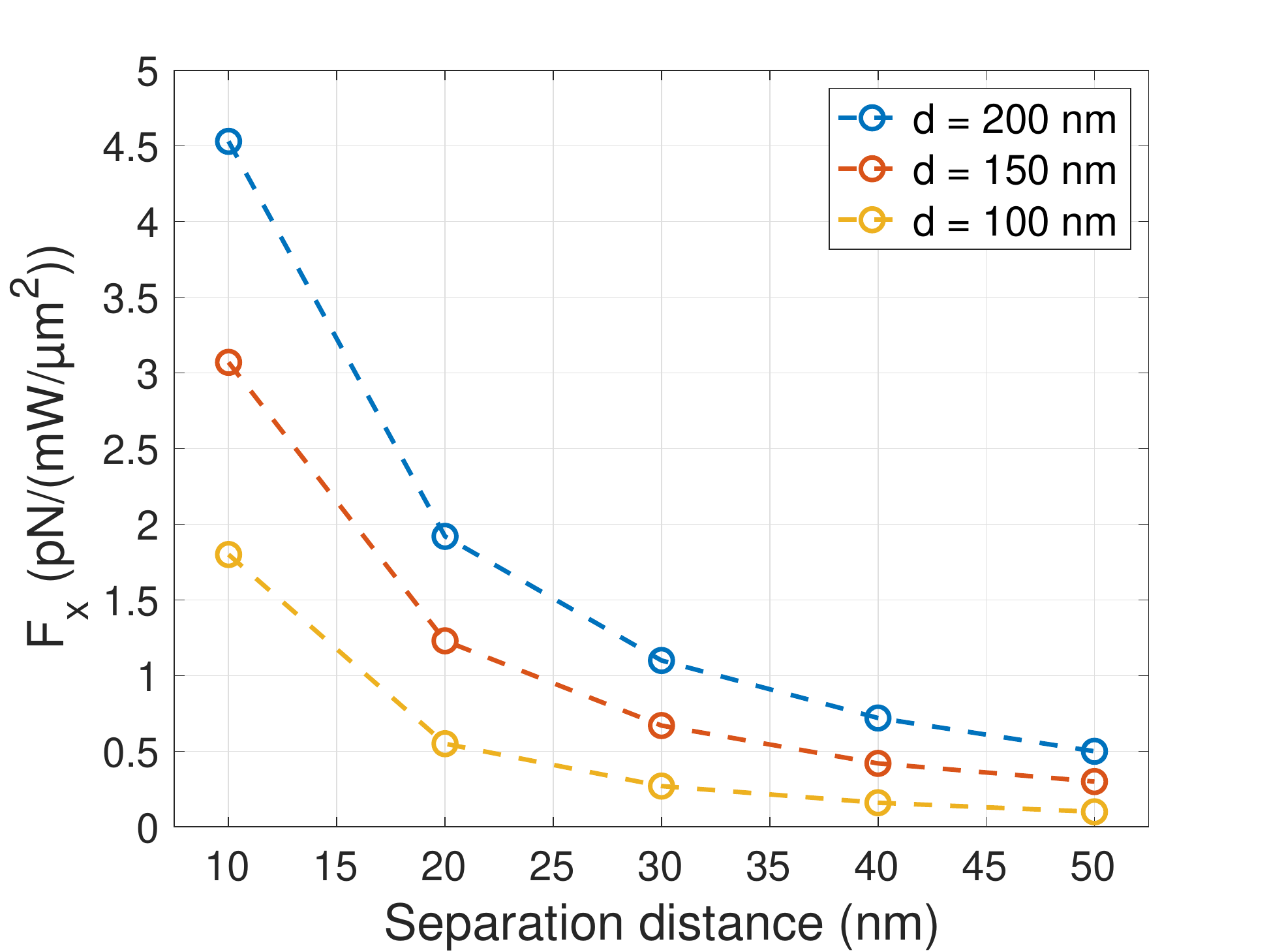}
    \caption{Optical forces in gold nanodisc dimers of diameter 100, 150 and 200~nm for Gaussian beam excitation. Nanodisc dimer with 100~nm diameter were excited at 625~nm, 150~nm diameter at 745~nm and 200~nm diameter at 900~nm respectively, with spot size equal to excitation wavelength for each case.}
    \label{fig:disc_gauss}
\end{figure}

Fig. \ref{fig:disc_gauss} shows the attractive optical forces in nanodisc dimers of radius 50, 75 and 100~nm respectively at varying inter-particle distances between 10 and 50~nm. Nanodisc dimers were excited using a Gaussian beam source where the excitation wavelength for each radius was chosen according to the plasmon resonance condition for the dimer at 10~nm gap between the discs. The 50~nm radius nanodiscs were excited at 625~nm, 75~nm radius nanodiscs at 745~nm and 100~nm nanodiscs at 900~nm respectively. The non-paraxial approximation was employed for generating the Gaussian beam, resulting in a beam width of 625~nm for exciting 50~nm nanodisc dimers, 745~nm beam width for exciting 75~nm radius nanodisc dimers and 900~nm beam width for exciting 100~nm radius nanodiscs. The magnitude of electric-field was 1~V/m for all cases of excitation.  

\subsection{Effect of rounding of the edges on scattering properties of nanocube dimer}
\begin{figure}
    \centering
    \includegraphics[width=0.485\textwidth]{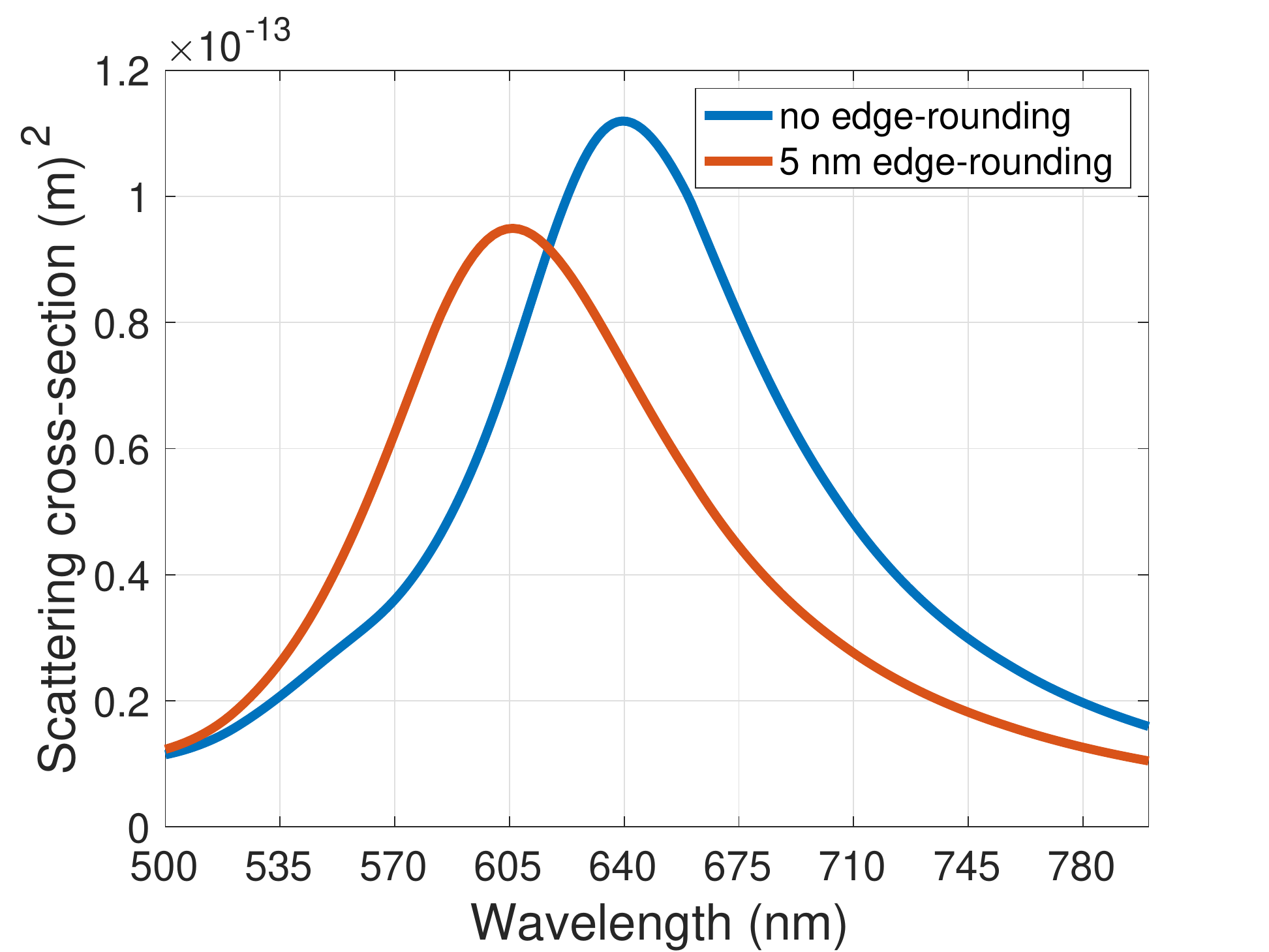}
    \caption{Comparison of scattering spectra of a gold nanocube dimer with and without rounding applied to its edges. nanocube dimer with 75~nm edge length and 10~nm gap was studied under plane wave excitation with edges rounded by a radius of 5~nm.} 
    \label{fig:edge_round}
\end{figure}

We observe the effect of the degree of sharpness of the edges of a nanocube dimer on its optical properties in Fig. \ref{fig:edge_round}. A blue-shift in the scattering spectra of a nanocube dimer of edge length 75~nm and 10~nm gap distance is observed upon rounding its edges by a radius of 5~nm.

\begin{figure*}[t]
    \begin{subfigure}{0.325\textwidth}
    \centering
    \includegraphics[width=\textwidth]{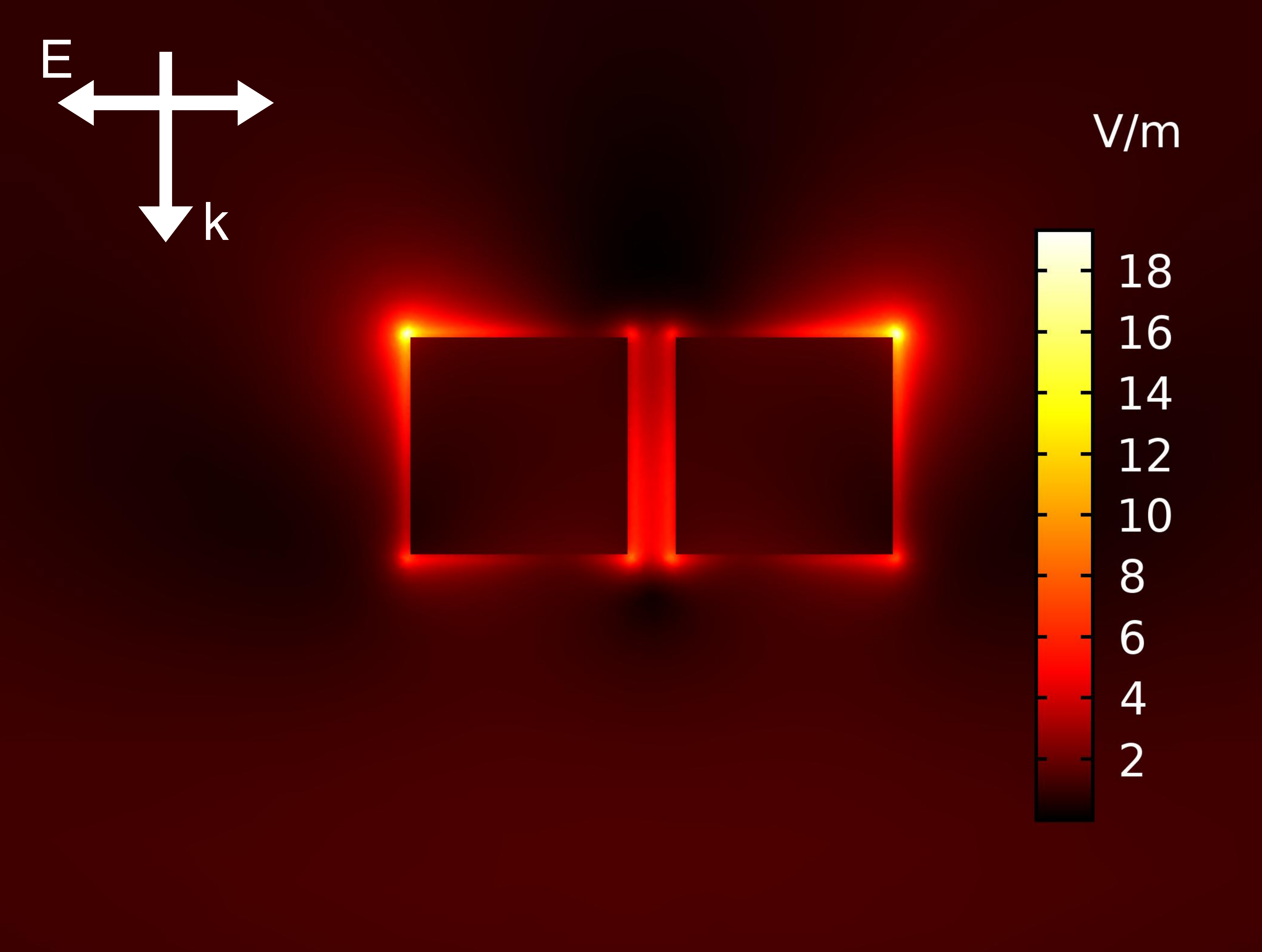}
    \caption{}
    \label{fig:100cube_604}
    \end{subfigure}
    \begin{subfigure}{0.325\textwidth}
    \centering
    \includegraphics[width=\textwidth]{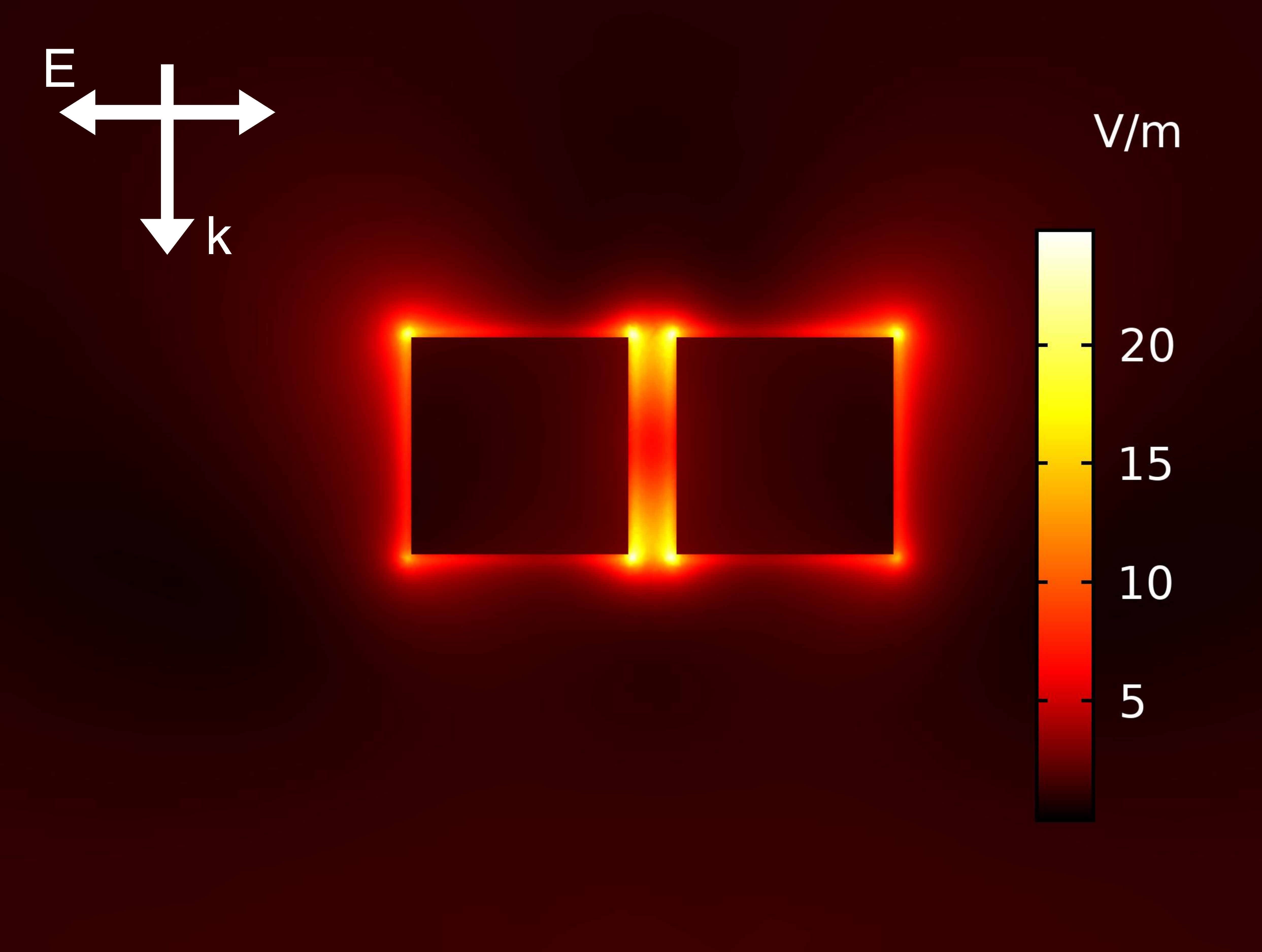}
    \caption{}
    \label{fig:100cube_670}
    \end{subfigure}
    \begin{subfigure}{0.325\textwidth}
    \centering
    \includegraphics[width=\textwidth]{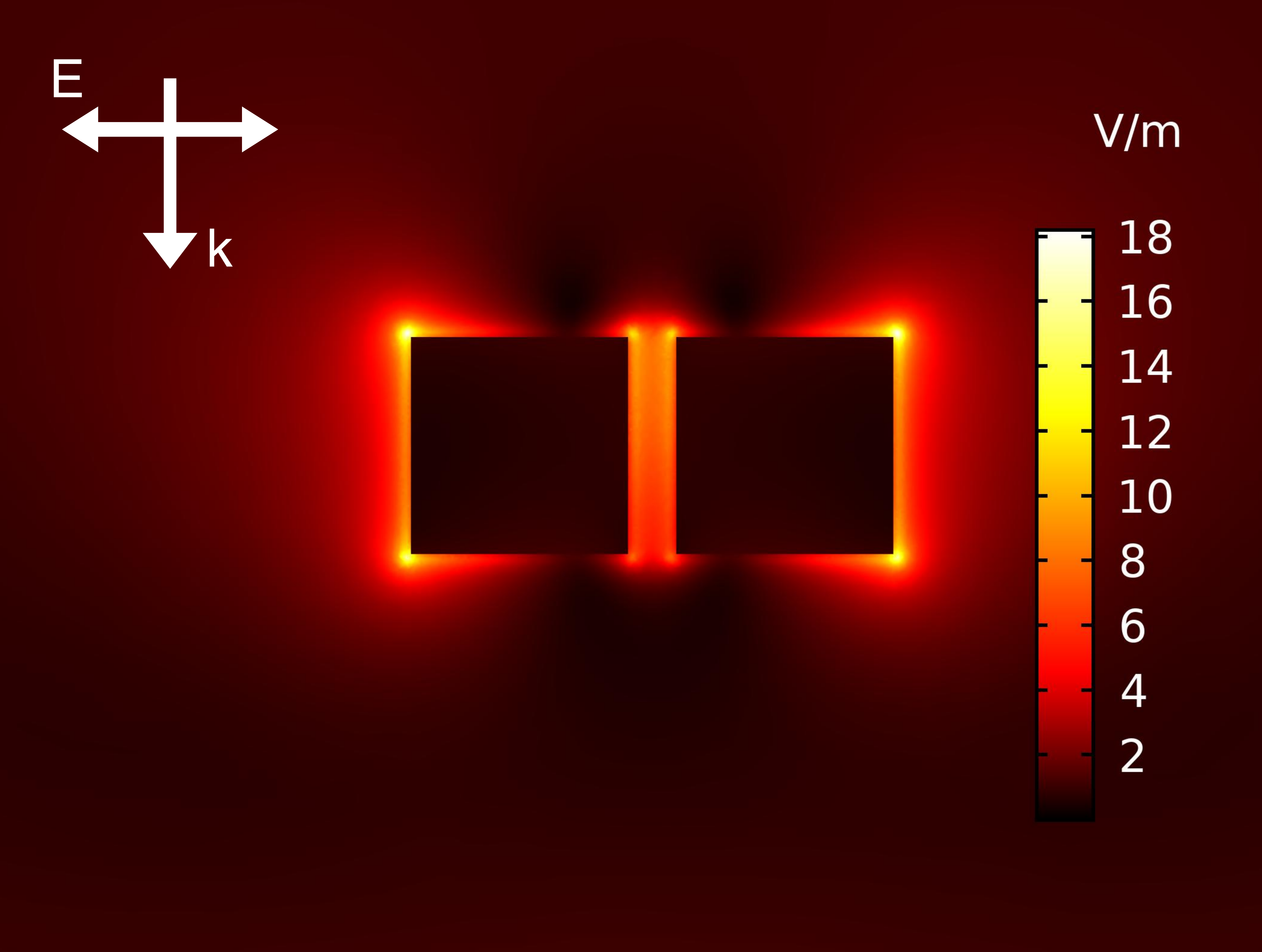}
    \caption{}
    \label{fig:100cube_800}
    \end{subfigure}
    \begin{subfigure}{0.325\textwidth}
    \centering
    \includegraphics[width=\textwidth]{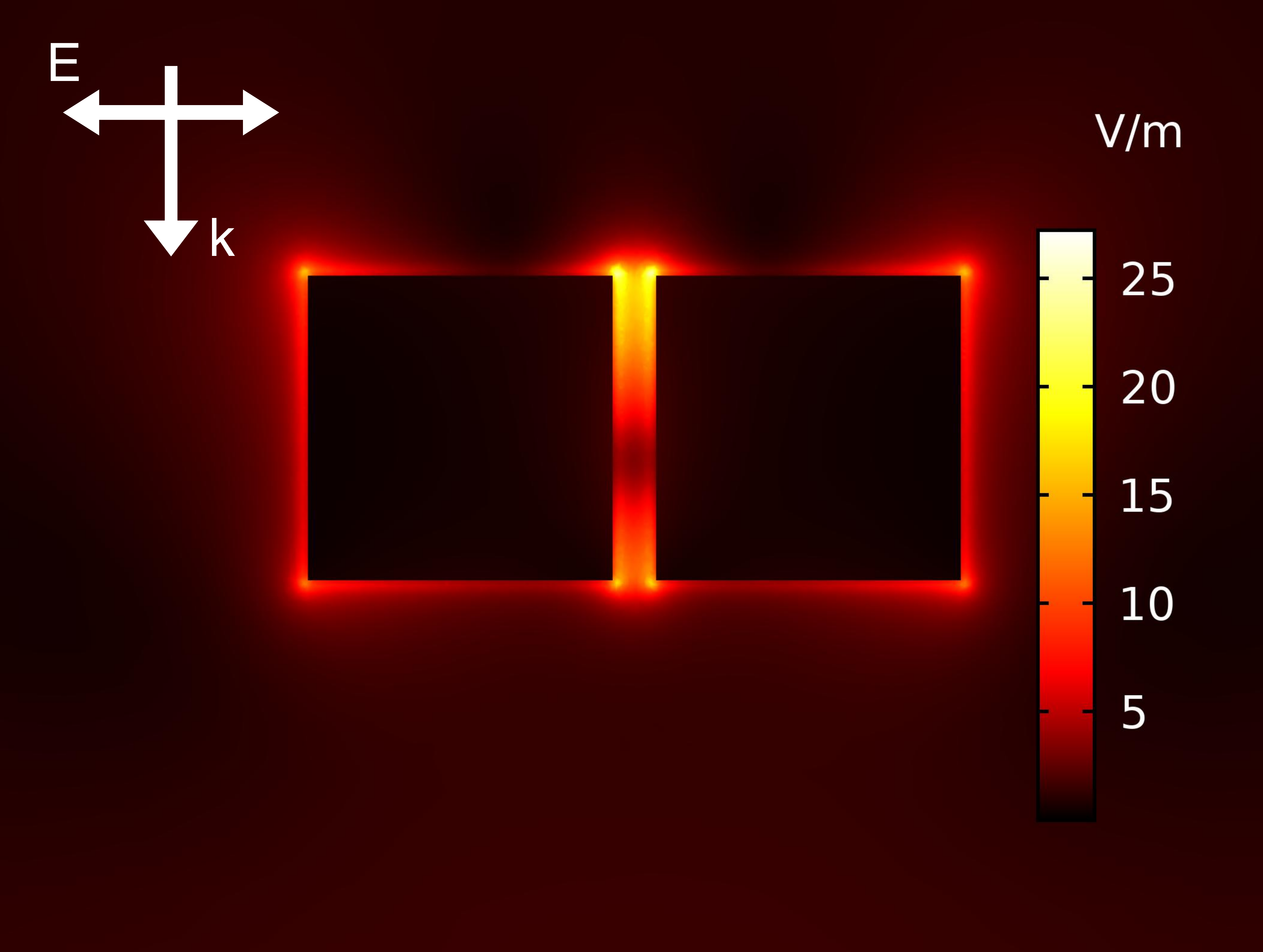}
    \caption{}
    \label{fig:150cube_880}
    \end{subfigure}
    \begin{subfigure}{0.325\textwidth}
    \centering
    \includegraphics[width=\textwidth]{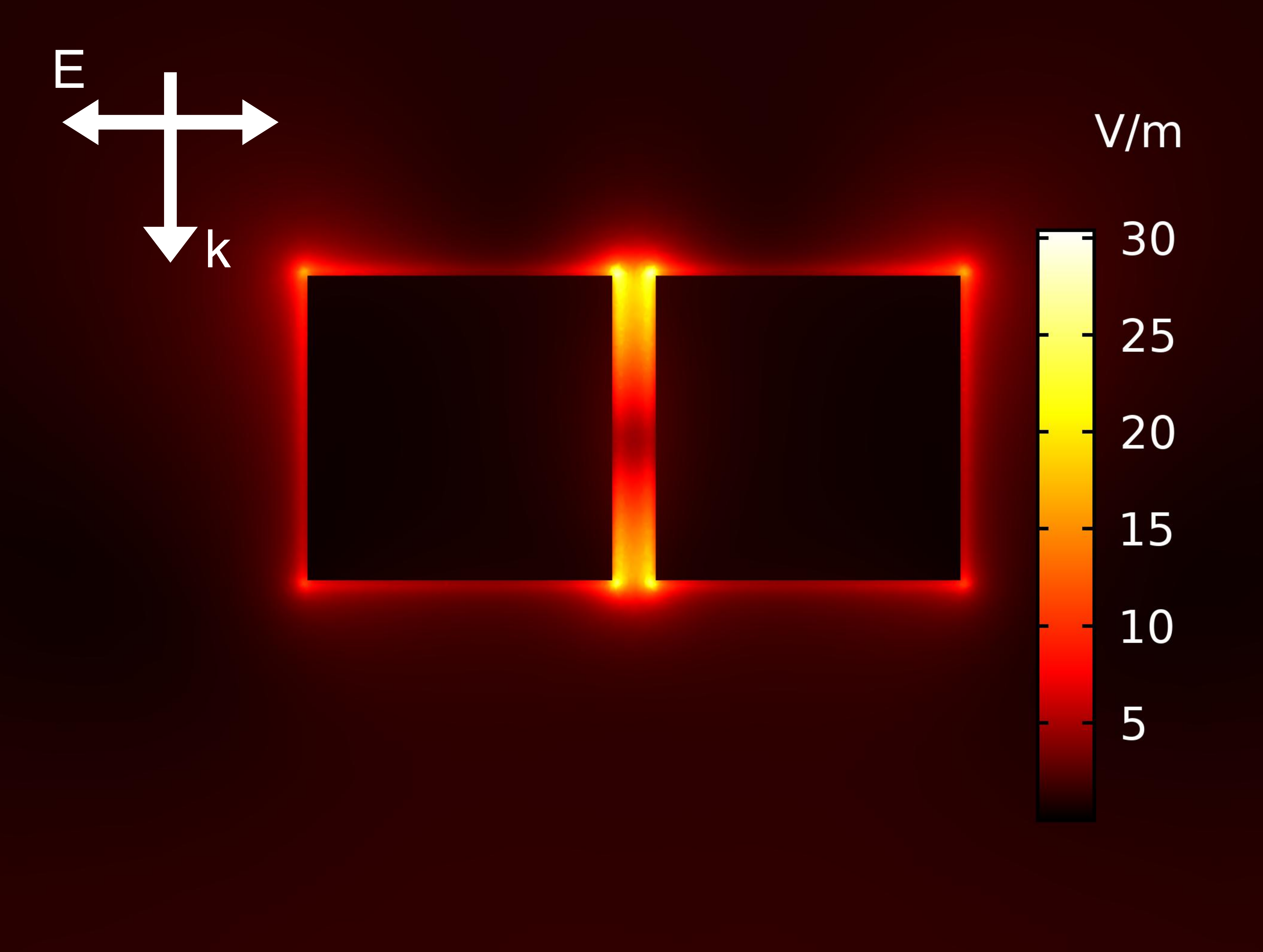}
    \caption{}
    \label{fig:150cube_862}
    \end{subfigure}
    \begin{subfigure}{0.325\textwidth}
    \centering
    \includegraphics[width=\textwidth]{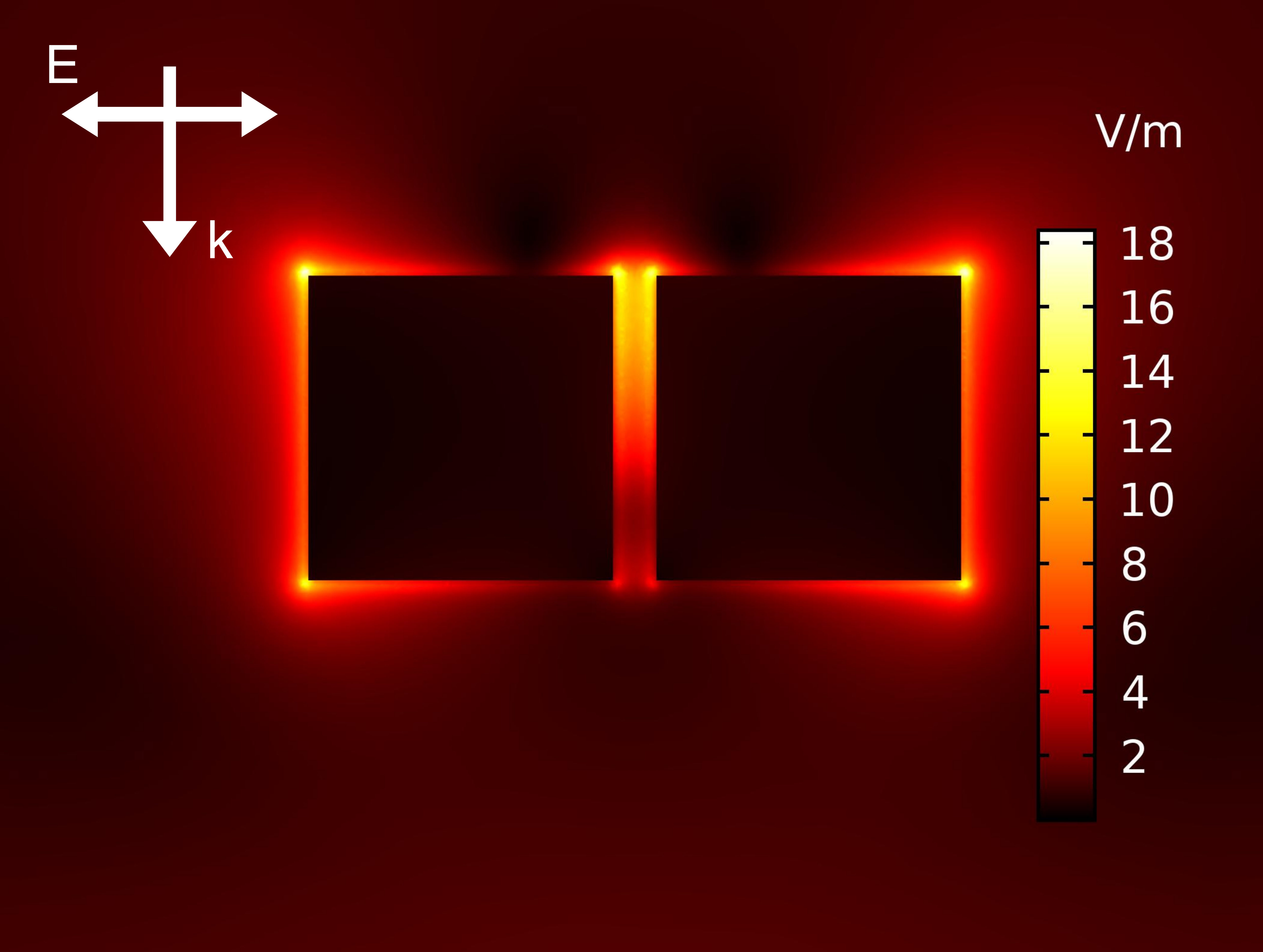}
    \caption{}
    \label{fig:150cube_936}
    \end{subfigure}
    \caption{x-z view of electric-field profile in gold nanocube dimers under plane wave excitation normal to the plane of nanocubes. Electric-field enhancement in 100~nm edge-length nanocube dimer with 10~nm gap distance at \textbf{a} 604~nm excitation, \textbf{b} at 670~nm excitation, \textbf{c} at 800~nm excitation. Electric-field enhancement in 150~nm edge-length nanocube dimer with 10~nm gap distance \textbf{d} at 880~nm excitation, \textbf{e} at 862~nm excitation, \textbf{f} at 936~nm excitation. The orientation of electric field is along the inter-particle axis of the dimer in x-direction.}
\end{figure*}

\subsection{Electric field enhancement in nanocube dimers}
Electric field (E-field) enhancement was calculated in nanocube dimer of edge length 100~nm and 150~nm at an inter-particle separation of 10~nm. The nanocube dimers were excited by plane wave travelling perpendicular to the plane of the cubes and the orientation of the electric field was kept along the inter-particle axis of the nanocubes for plasmonic coupling between the nanocubes. Fig. \ref{fig:100cube_604}, \ref{fig:100cube_670} and \ref{fig:100cube_800} show the E-field enhancement profile in nanocube dimers with 100~nm edge length at 604~nm, 670~nm and 800~nm respectively. At resonant excitation condition (670~nm), a strong localisation of the electric field is seen in the gap region between the nanocubes, also referred to as the hot-spot for local field enhancement. When the excitation source wavelength is red-shifted w.r.t the plasmon resonance to 800~nm, the regions of localisation of the E-field shift position towards outer edges of the nanocubes. In case of excitation wavelength of 604~nm, which is blue-shifted to the resonance, the localisation can be seen present at outer corners of the nanocube dimer.  

Fig. \ref{fig:150cube_880}, \ref{fig:150cube_862} and \ref{fig:150cube_936} show the electric-field enhancement profile in nanocube dimer of edge-length 150~nm at 880~nm, 862~nm and 936~nm respectively. Similar to 100~nm edge length nanocube dimer, the localisation of the E-field at resonant excitation (862~nm) is observed in the gap region between the two nanocubes. On changing the excitation source wavelength to 880~nm, the localisation becomes asymmetric in the gap region. Upon red-shifting the excitation wavelength further to 936~nm, the localisation of the E-field shifted position towards the outer edges of the nano-dimer. Compared to nanocubes with 100~nm edge length, greater enhancement of E-field was observed in nanocubes with 150~nm edge length.

\subsection{Spectral response of optical forces in nanodisc, nanocube and nanobar dimers}
\begin{figure*}
\begin{subfigure}[h]{0.40\textwidth}
\centering
\includegraphics[width=\textwidth]{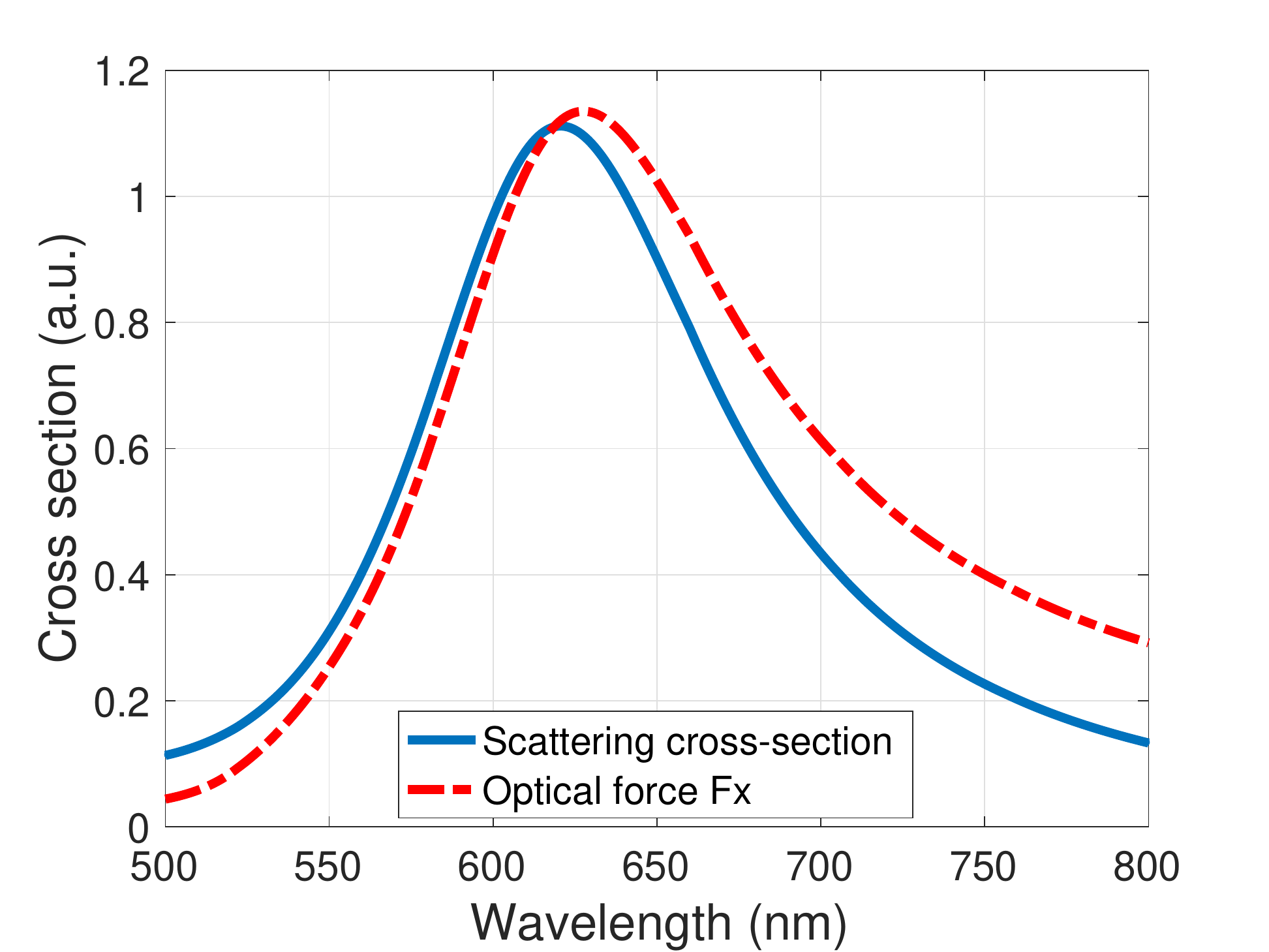}
\caption{}
\label{fig:disc50_opt}
\end{subfigure}    
\begin{subfigure}[h]{0.40\textwidth}
\centering
\includegraphics[width=\textwidth]{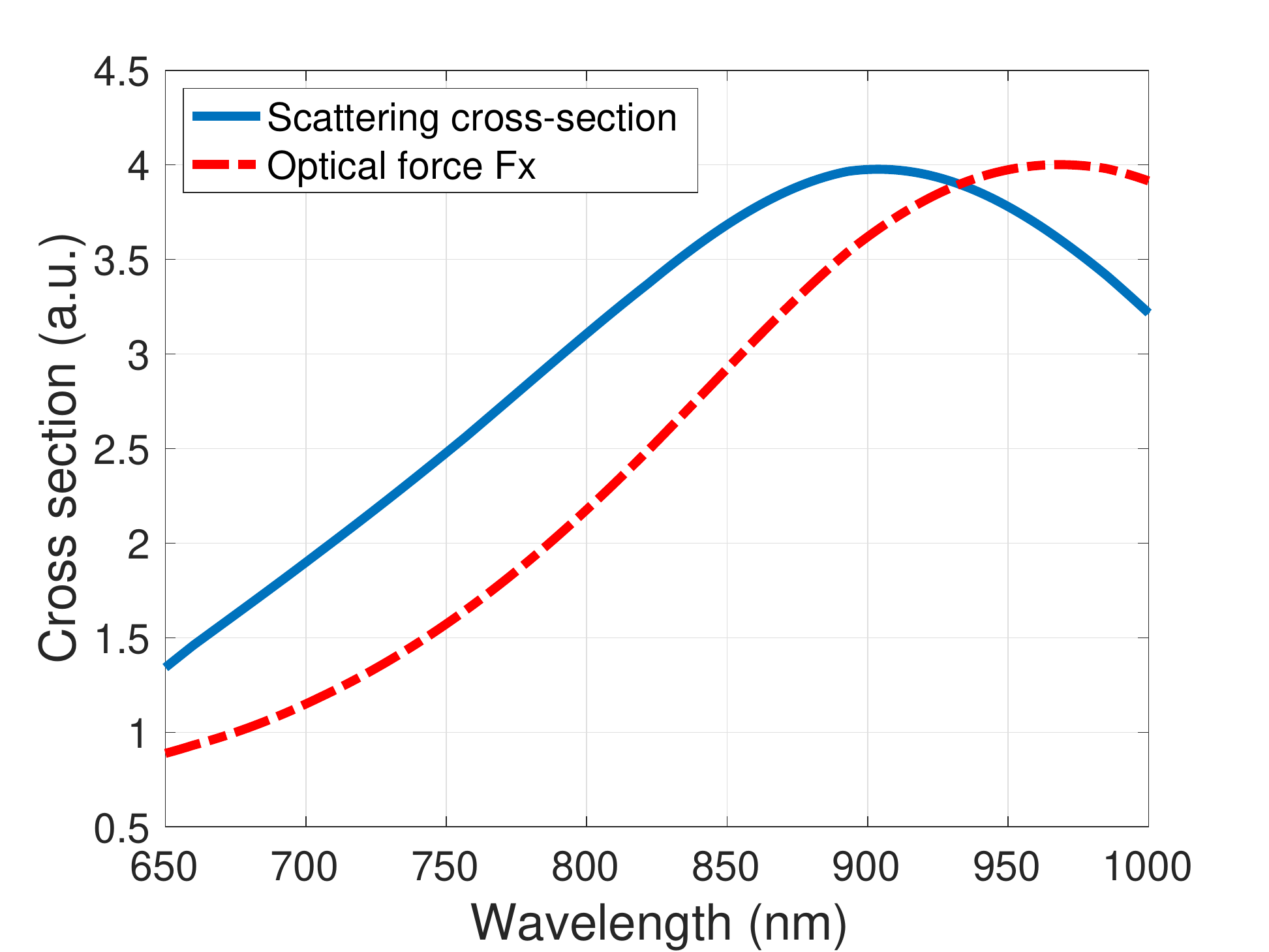}
\caption{}
\label{fig:disc100_opt}
\end{subfigure}
\caption{Spectral response of optical forces in nanodisc dimers. \textbf{a} Response of optical forces with wavelength in nanodisc dimers with 50~nm radius and 10~nm gap distance. \textbf{b} Response of optical forces with wavelength in nanodisc dimers with 100~nm radius and 10~nm gap distance.}
\end{figure*}

\begin{figure*}
\begin{subfigure}[h]{0.40\textwidth}
\centering
\includegraphics[width=\textwidth]{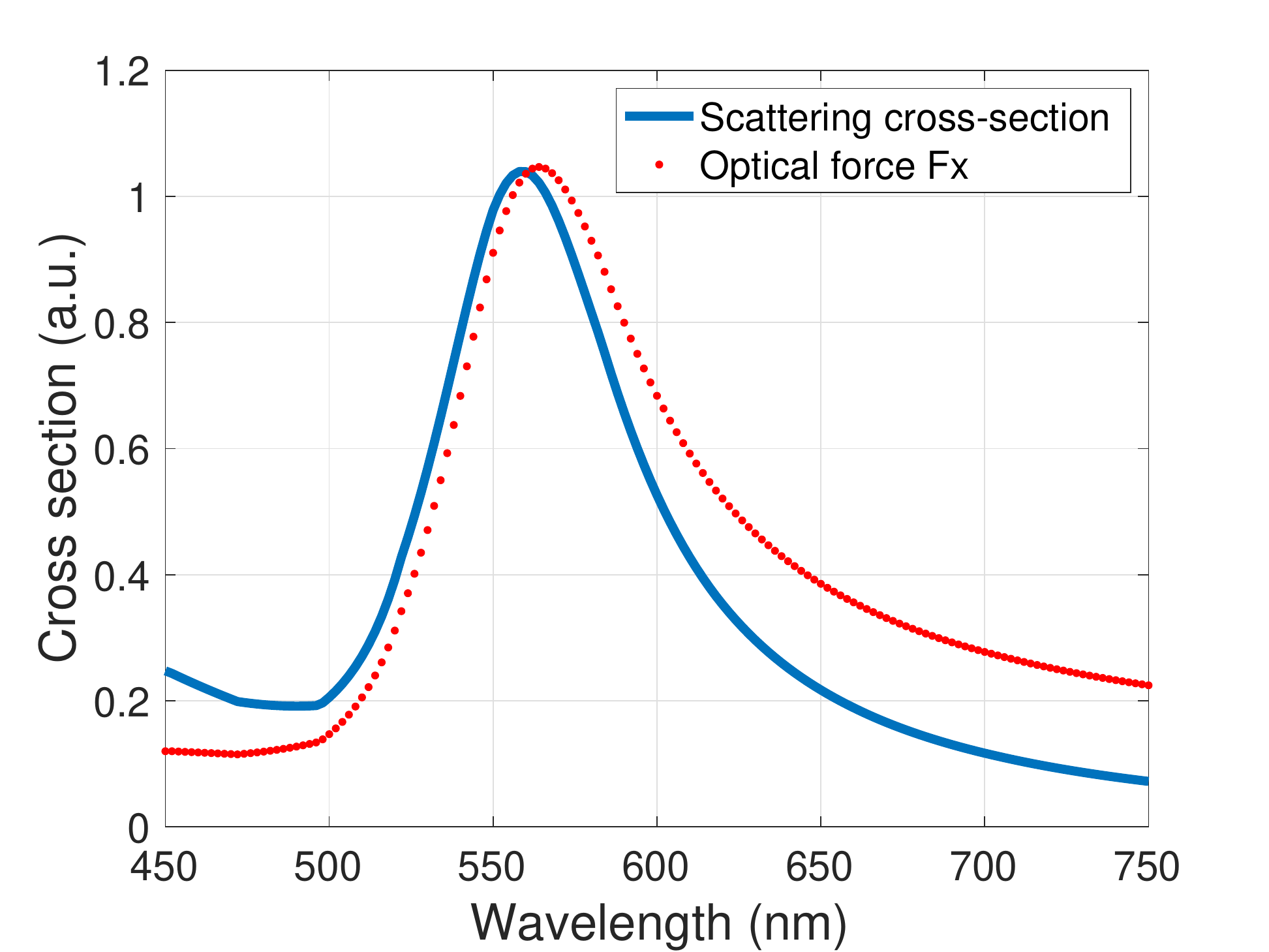}
\caption{}
\label{fig:cube50_opt}
\end{subfigure}
\begin{subfigure}[h]{0.40\textwidth}
\centering
\includegraphics[width=\textwidth]{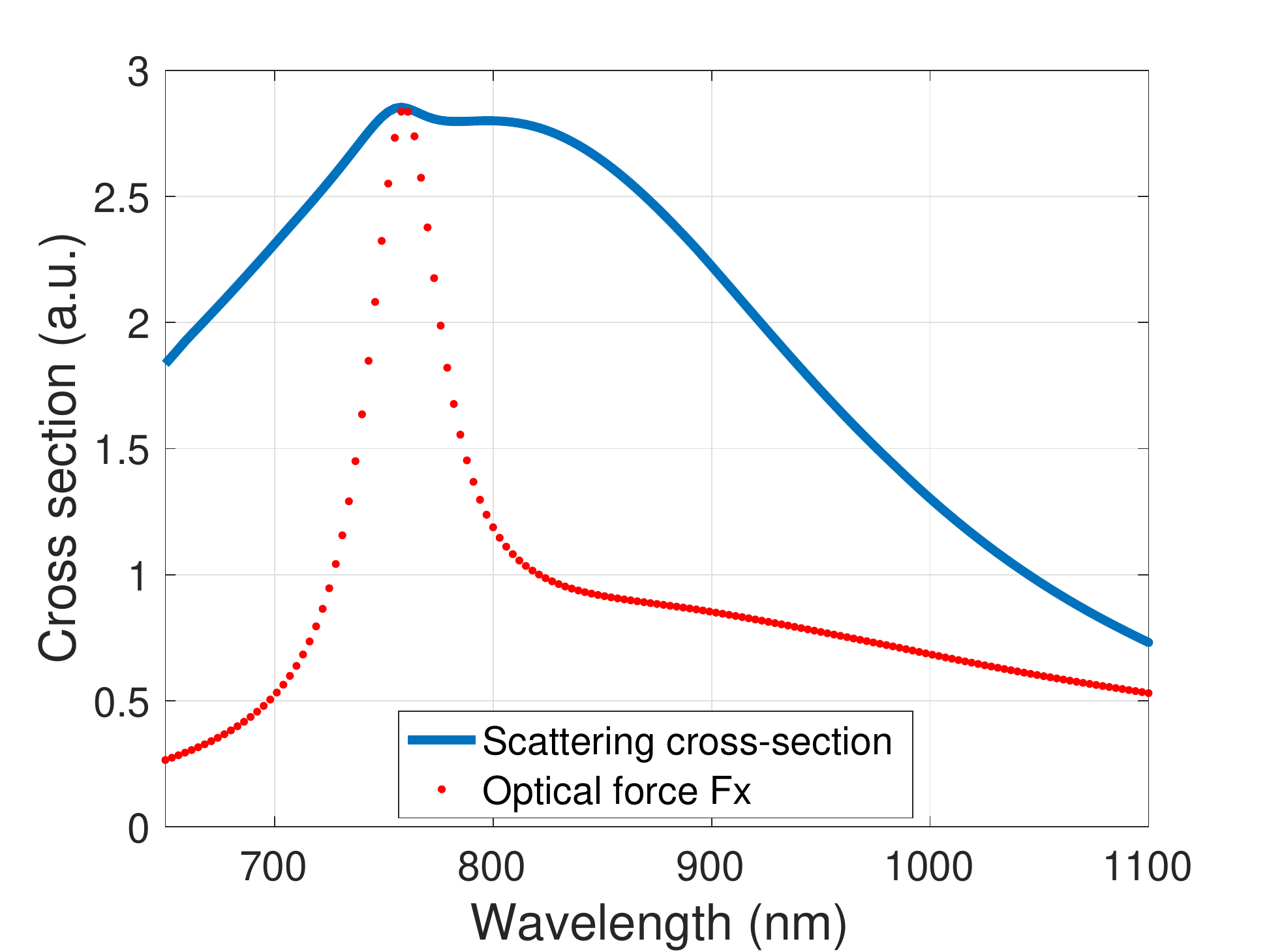}
\caption{}
\label{fig:cube125_opt}
\end{subfigure}
\caption{Spectral response of optical forces in nanocube dimers. \textbf{a} Response of optical forces with wavelength in nanocube dimers with 50~nm edge-length and 10~nm gap distance. \textbf{b} Response of optical forces with wavelength in nanocube dimers with 125~nm edge-length and 10~nm gap distance.}
\end{figure*}

\begin{figure*}[!]
\begin{subfigure}[h]{0.40\textwidth}
\centering
\includegraphics[width=\textwidth]{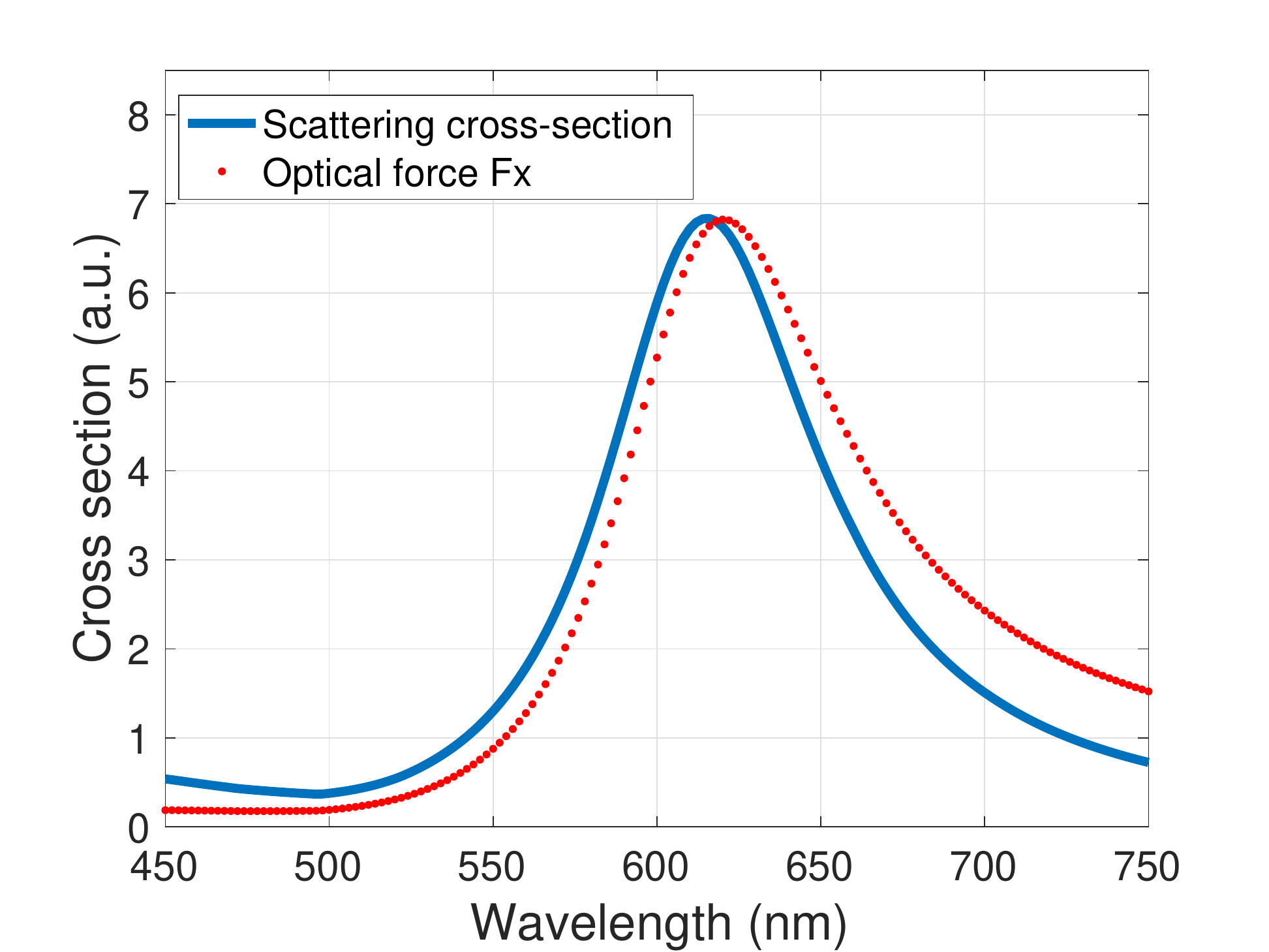}
\caption{}
\label{fig:bar50_opt}
\end{subfigure}
\begin{subfigure}[h]{0.40\textwidth}
\centering
\includegraphics[width=\textwidth]{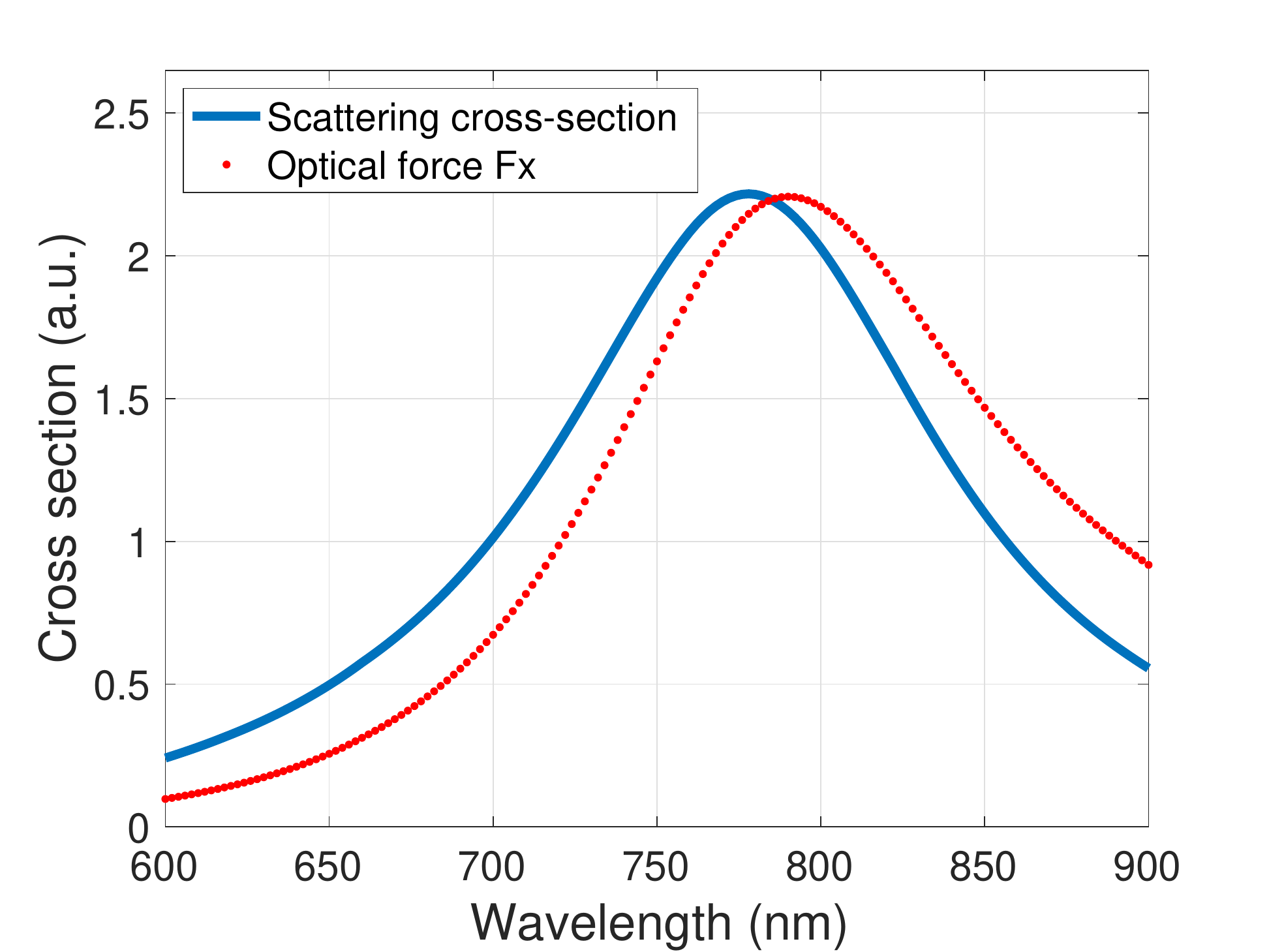}
\caption{}
\label{fig:bar125_opt}
\end{subfigure}
\caption{Spectral response of optical forces in nanobar dimers. \textbf{a} Response of optical forces with wavelength in nanobar dimers with 50~nm side-length and 10~nm gap distance. \textbf{b} Response of optical forces with wavelength in nanobar dimer with 125~nm side-length and 10~nm gap distance.}
\end{figure*}

The relation of optical forces with wavelength was studied in nanodiscs, nanocubes and nanobars at an inter-particle distance of 10~nm. The scattering properties of the respective geometries were plotted to observe the correlation between the optical properties and the internal plasmonic forces present in them. The optical force spectra was fitted  1:1 in magnitude to the scattering spectra for gaining better visibility into the wavelength dependent trends exhibited by the optical forces compared to the scattering properties for each dimer case. As we can see in Fig. \ref{fig:disc50_opt}, the spectra of attractive optical forces between the 50~nm radius nanodiscs exhibits a behaviour with wavelength which is close to the scattering properties of the dimer. When the diameter of the nanodiscs is changed to 100~nm, we observe broadening of the spectrum associated with the optical forces and scattering Fig. \ref{fig:disc100_opt}. For nanocube dimer with 50~nm edge-length, we observe that the variation of attractive optical forces with wavelength display a trend similar to the scattering properties of the dimer Fig. \ref{fig:cube50_opt}. On increasing the edge-length of the cube to 125~nm, we see distinct behaviour of the optical force from the scattering spectra. As we can see in Fig. \ref{fig:cube125_opt} the optical force spectra narrows for the dimer while the scattering spectra of the dimer further broadens and exhibits an additional peak at 809~nm. For nanobar dimer with thickness 50~nm, side-length 50~nm and length 75~nm, we see that the attractive optical forces follow the scattering spectra closely, as shown in Fig. \ref{fig:bar50_opt}. When increasing the length of the nanobar to 125~nm, we see broadening of the spectra associated with both the optical forces and scattering Fig. \ref{fig:bar125_opt}. The spectra of optical forces red-shifts w.r.t the scattering spectra on increasing the length of nanobars, which is also observed in case of nanodiscs on increasing their radius. The response observed in nanocubes however is distinct from the other two discussed geometries.  

\begin{figure*}
\begin{subfigure}[h]{0.325\textwidth}
\centering
\includegraphics[width=\textwidth]{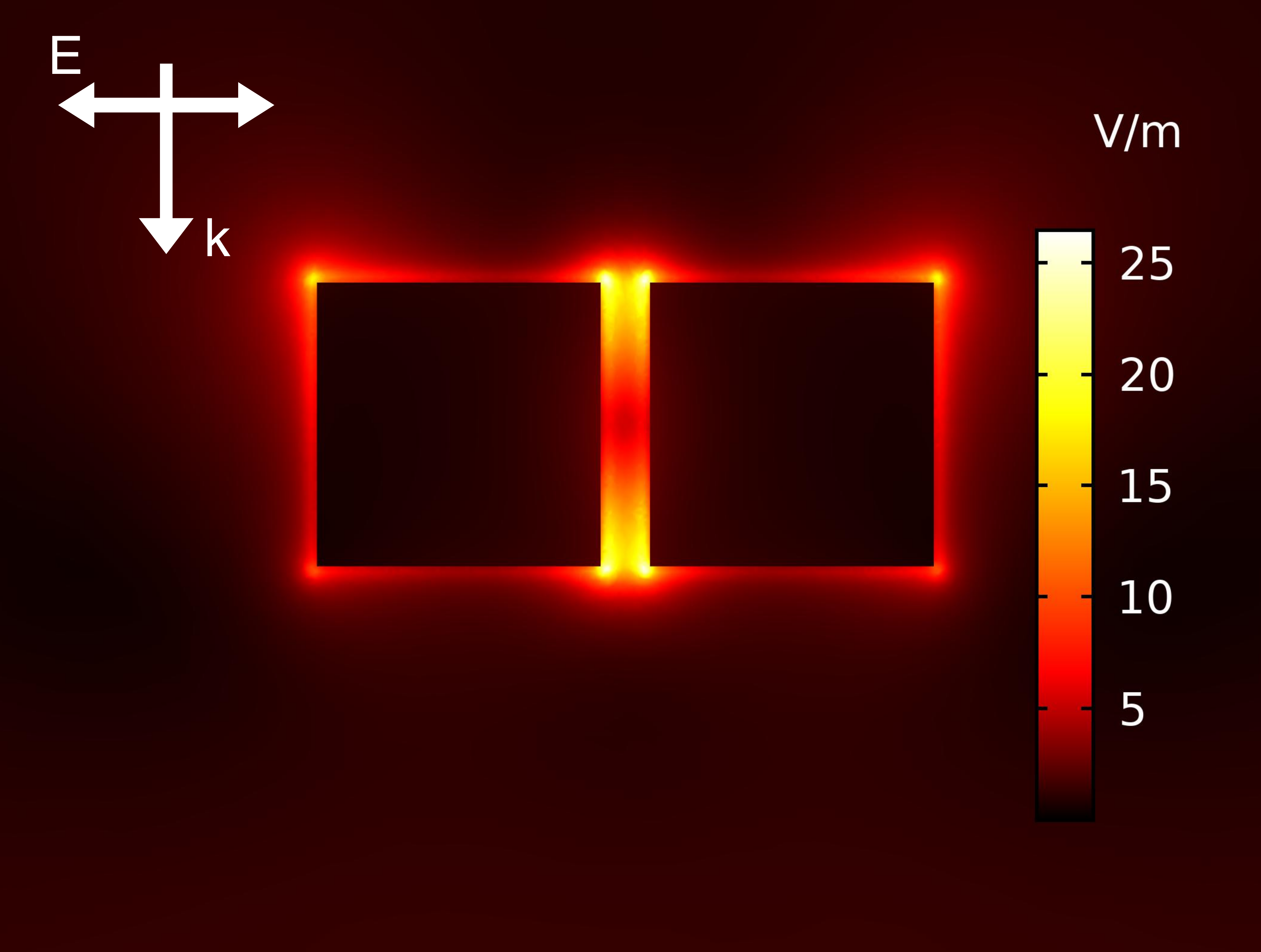}
\caption{}
\label{fig:125cube_755}
\end{subfigure}
\begin{subfigure}[h]{0.325\textwidth}
\centering
\includegraphics[width=\textwidth]{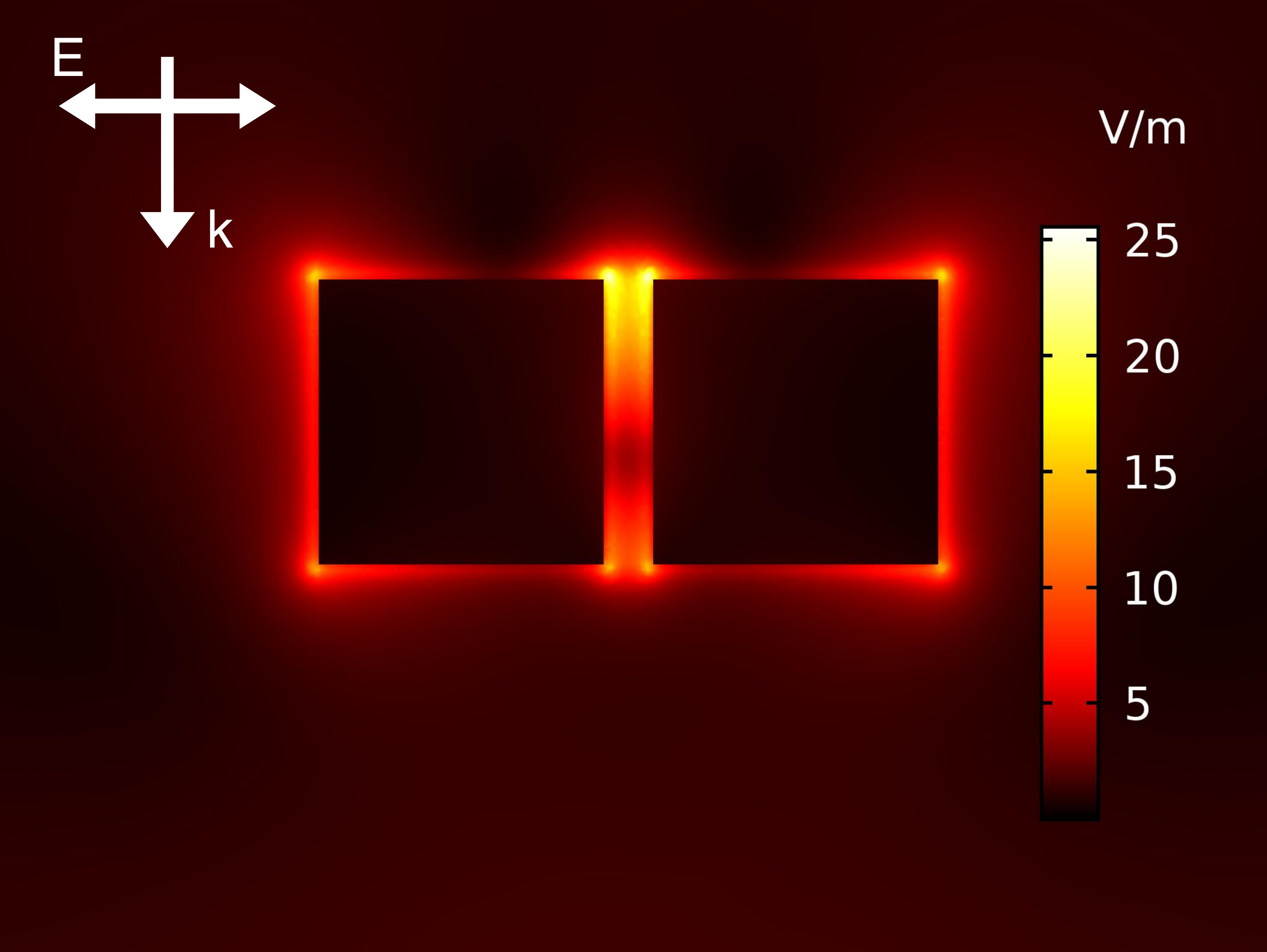}
\caption{}
\label{fig:125cube_779}
\end{subfigure}
\begin{subfigure}[h]{0.325\textwidth}
\centering
\includegraphics[width=\textwidth]{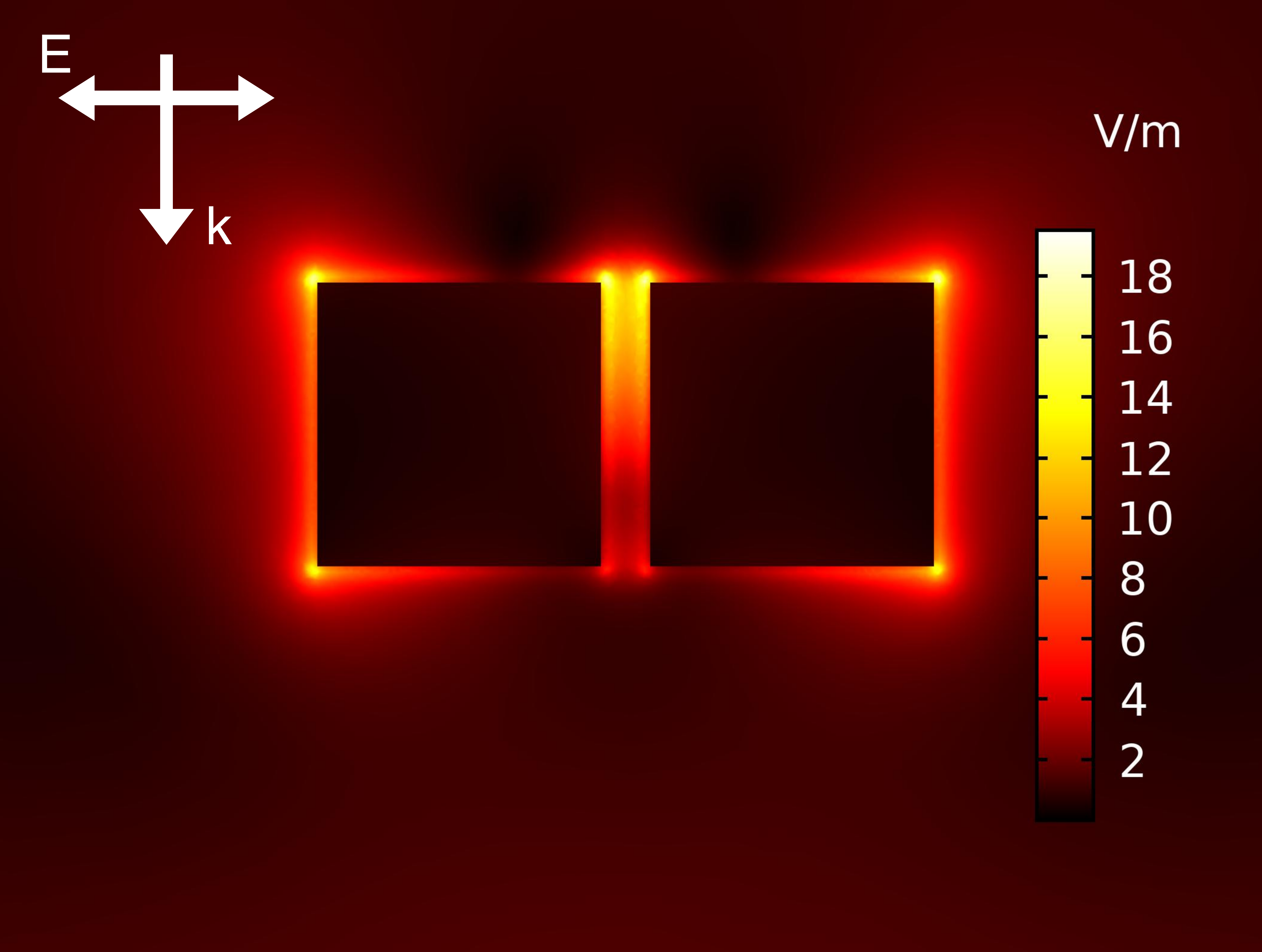}
\caption{}
\label{fig:125cube_809}
\end{subfigure}
\begin{subfigure}[h]{0.325\textwidth}
\centering
\includegraphics[width=\textwidth]{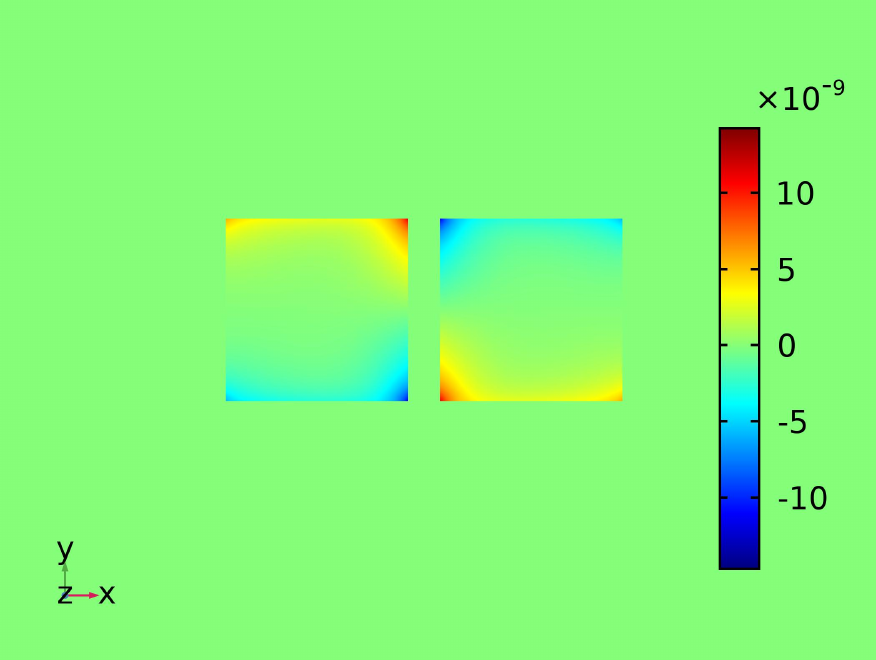}
\caption{}
\label{fig:pol125_755}
\end{subfigure}
\begin{subfigure}[h]{0.325\textwidth}
\centering
\includegraphics[width=\textwidth]{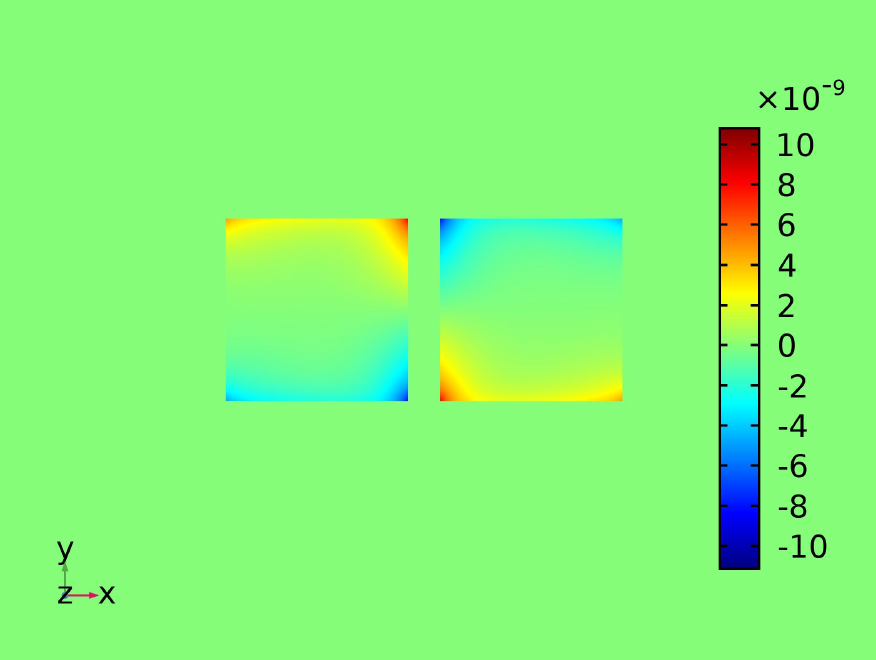}
\caption{}
\label{fig:pol125_779}
\end{subfigure}
\begin{subfigure}[h]{0.325\textwidth}
\centering
\includegraphics[width=\textwidth]{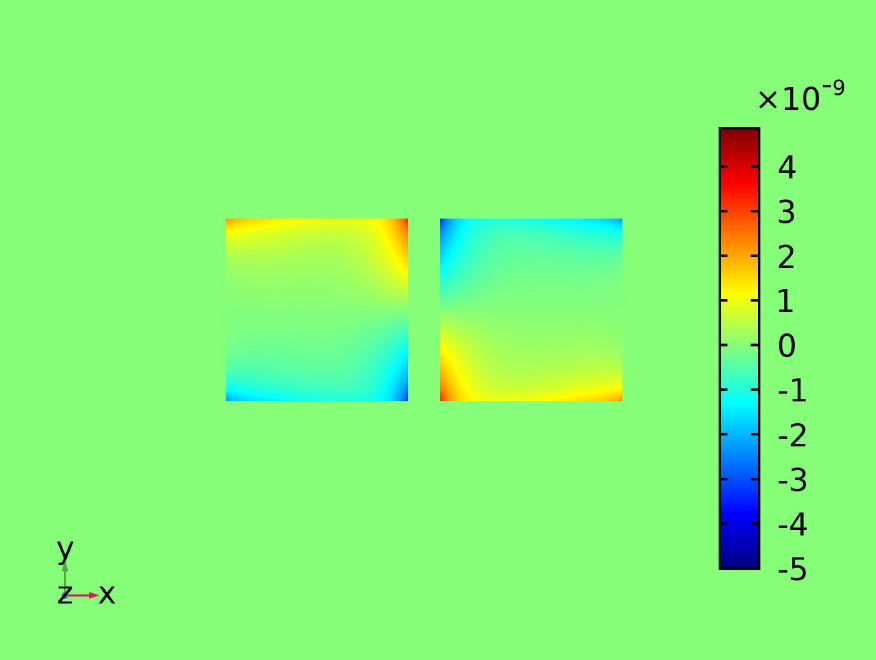}
\caption{}
\label{fig:pol125_809}
\end{subfigure}
\caption{Correlation of electric field enhancement with polarisation density in nanocube dimers under plane wave excitation normal to the plane of the nanocubes. x-z view of electric-field enhancement in 125~nm edge-length nanocube dimer with 10~nm gap at \textbf{a} 755~nm excitation, \textbf{b} 779~nm excitation, \textbf{c} 809~nm excitation. Polarisation density for 125~nm edge-length nanocube dimer with 10~nm gap at \textbf{d} 755~nm excitation, \textbf{e} 779~nm excitation, \textbf{f} 809~nm excitation. The electric field was aligned along the inter-particle axis of the dimer in x-direction.}
\end{figure*}

\subsection{Polarisation density map of nanocube dimer}
The correlation of polarisation density with electric field enhancement was studied in nanocube dimer with 125~nm edge-length and 10~nm inter-particle spacing. Incident light was injected along z-axis with electric-field vector aligned along the inter-particle axis of the cube dimer, i.e x-axis. Fig. \ref{fig:125cube_755} shows the E-field enhancement in the cube dimer at resonant excitation conditions 755~nm. The calculation of the polarisation density corresponding to resonant excitation for the dimer at 755~nm shows a magnitude of $14 \times 10^{-9}$~C/m$^2$ as shown in Fig. \ref{fig:pol125_755}. The polarisation density drops in magnitude to $10.3 \times 10^{-9}$~C/m$^2$ at 779~nm excitation wavelength \ref{fig:pol125_779}. The change in polarisability causes an asymmetry in local-field enhancement along the gap region of the dimer. As it can be seen in Fig. \ref{fig:125cube_779}, the hot-spot for field enhancement becomes more prominent around the upper corner of the nanocube dimer. When the excitation wavelength is red-shifted to 809~nm, corresponding to the second peak in the scattering spectra of the nanodimer \ref{fig:cube125_opt}, the hot-spots of electric-field enhancement shift towards the outer edges of the nanocube dimer, as shown in Fig. \ref{fig:125cube_809}. The associated polarisation density for the 809~nm excitation reveals a drop in its magnitude to $4.5 \times 10^{-9}$~C/m$^2$. These observations convincingly show the possibility to tune the electric-field enhancement in nanocube dimers by changing the magnitude of polarisation density exhibited by them, which can be controlled by wavelength of the excitation source.  

\begin{figure*}
\begin{subfigure}[h]{0.40\textwidth}
\centering
\includegraphics[width=\textwidth]{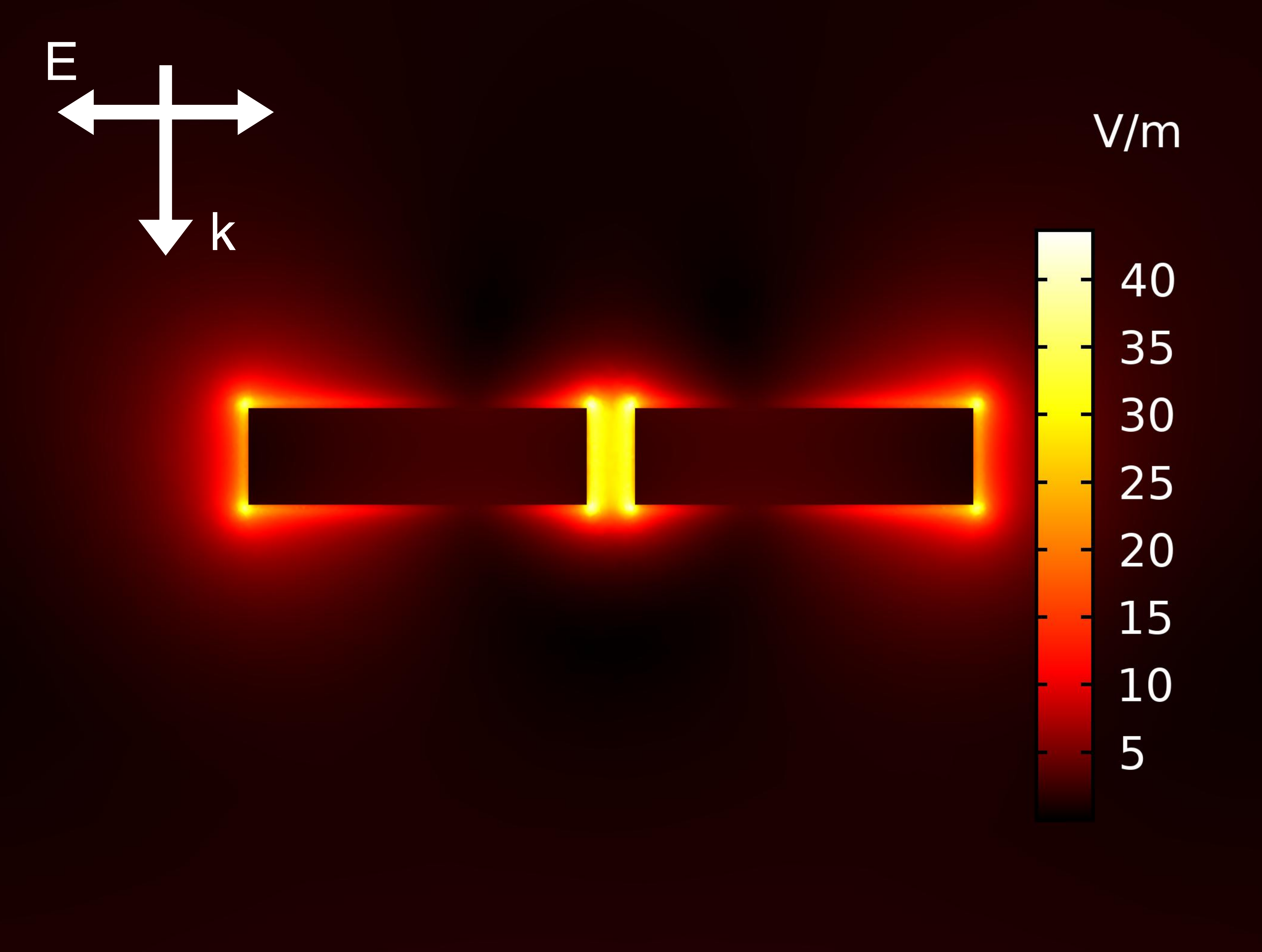}
\caption{}
\label{fig:150bar_10d}
\end{subfigure}
\begin{subfigure}[h]{0.40\textwidth}
\centering
\includegraphics[width=\textwidth]{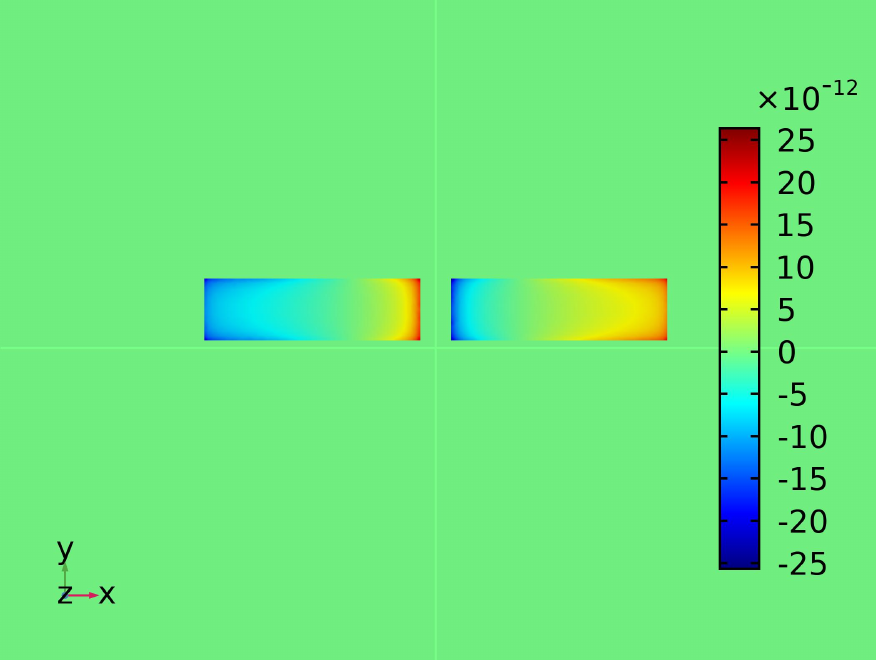}
\caption{}
\label{fig:150bar_pol}
\end{subfigure}
\caption{x-z view of electric field enhancement and polarisation density in nanobar dimer under plane wave excitation normal to the plane of nanobars. \textbf{a} Local field enhancement and \textbf{b} polarisation density of nanobar dimer at 795~nm excitation. The nanobar dimer had thickness and side length 50~nm and 150~nm length with 10~nm gap.}
\end{figure*}

\subsection{Electric field enhancement and polarisation density map of nanobar dimer}
Electric field enhancement and polarisation density were studied under plane wave excitation in nanobar dimer of side length and thickness of 50~nm and a longitudinal length of 150~nm with 10~nm inter-particle separation distance. Light was perpendicularly incident on the plane of the nanobars, and the electric field vector was aligned along the inter-particle axis of the dimer. For resonant excitation condition at 795~nm, we observe a strong enhancement of electric-field in the gap region between the nanobars, as shown in Fig. \ref{fig:150bar_10d}. The corresponding polarisation density for the dimer at 795~nm excitation reveals a magnitude of $26 \times 10^{-12}$~C/m$^2$. The polarisation map shows localisation of opposite charges at the facing sides of the dimer, as shown in Fig. \ref{fig:150bar_pol}, with a symmetric distribution of charge across the nanobar dimer's surface (along the y-axis).

\begin{figure*}
\begin{subfigure}[h]{0.40\textwidth}
\centering
\includegraphics[width=\textwidth]{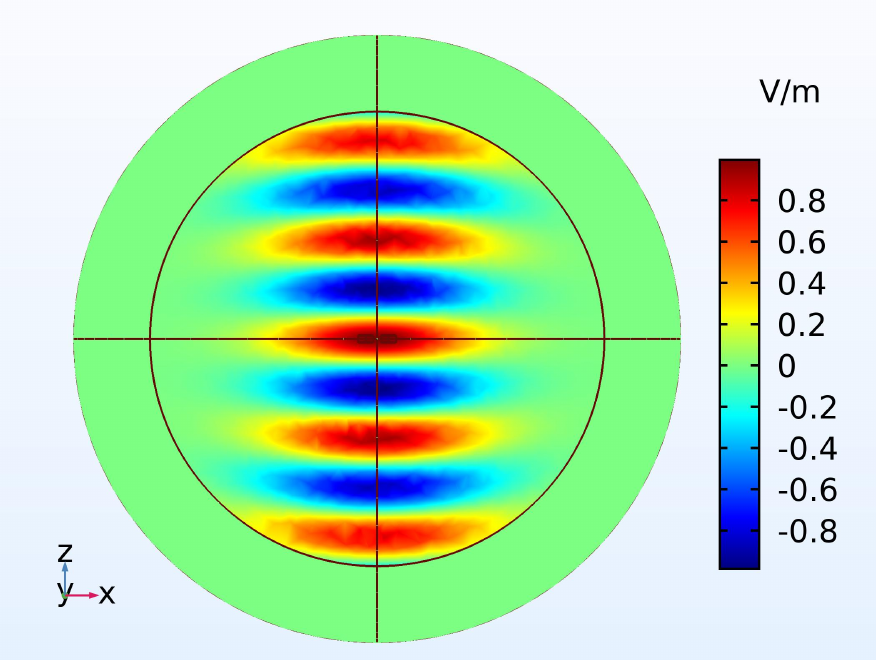}
\caption{}
\label{fig:gauss1}
\end{subfigure}
\begin{subfigure}[h]{0.40\textwidth}
\centering
\includegraphics[width=\textwidth]{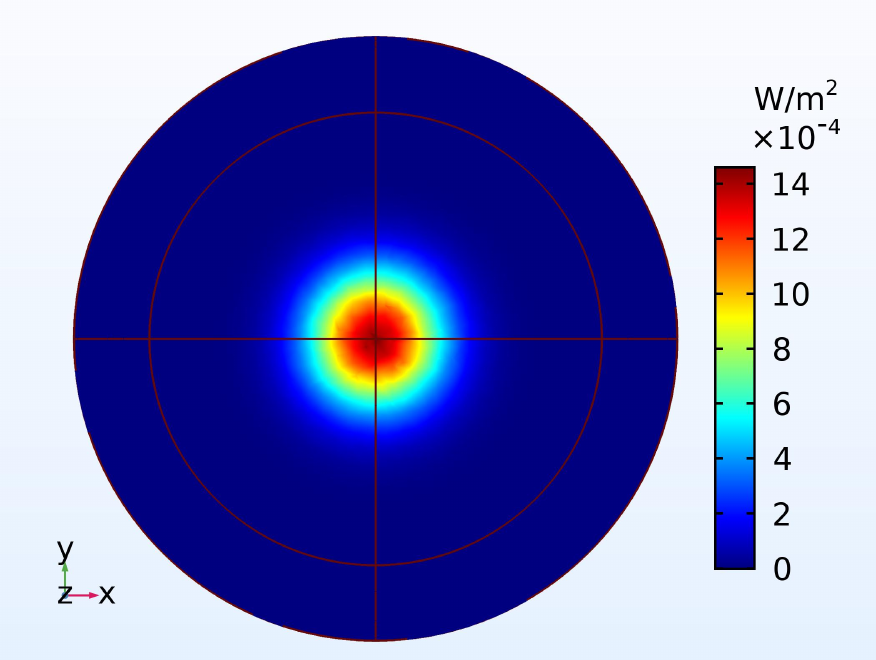}
\caption{}
\label{fig:gauss2}
\end{subfigure}
\caption{Electric field plots of non-paraxial Gaussian beam used as excitation source. \textbf{a} Electric field distribution for a propagating 3D gaussian beam. \textbf{b} Intensity distribution for the Gaussian beam in the x-y plane.}
\end{figure*}

\subsection{Simulation methodology for generating Gaussian beam source}

The Gaussian beam source was formulated using COMSOL Multiphysics (Version 5.6) using the non-paraxial approximation \footnote{https://www.comsol.com/blogs/the-nonparaxial-gaussian-beam-formula-for-simulating-wave-optics/}. The formulation works on the idea of constructing a beam by addition of waves travelling in multiple directions with the same wave number k. The resulting wave allows the possibility for realising a beam with a waist size which is comparable to its wavelength i.e 1:1. This is however not achievable using the conventional paraxial approximation, since reducing the waist size of the beam results in angular deviation in the beam propagation direction. Fig. \ref{fig:gauss1}  shows the side view of a Gaussian beam propagating along the z-axis. The nanoparticle dimer was placed in the focus of the beam at the central position. The alternatively varying electric and magnetic fields are identified from their respective colors on the intensity scale. Fig. \ref{fig:gauss2} shows the intensity distribution of the Gaussian beam in the plane of the nanoparticles i.e. x-y plane. It could be seen that the electro-magnetic field intensity of the beam W/m$^2$ shows a Gaussian distribution, with maximum intensity localised at the center of the beam which falls exponentially to the outside from all directions.

\end{document}